\documentclass[10pt,aps,prb,twocolumn,english,amsmath,amssymb,floatfix,superscriptaddress,longbibliography,preprintnumbers]{revtex4-2}

\usepackage{graphicx}
\usepackage[dvipsnames]{xcolor}
\usepackage{amsmath,braket}

\usepackage[normalem]{ulem}

\makeatother

\usepackage{babel}
\usepackage{bm}

\usepackage[colorlinks,urlcolor=orange,citecolor=blue,linkcolor=magenta]{hyperref}

\begin{document}
\title{Quantum Hall effect in three-dimensional lattice induced by\\ Wannier-Stark-Landau localization}
\author{D.-H.-Minh Nguyen}
\email{d.h.minh.ng@gmail.com}
\affiliation{Donostia International Physics Center, 20018 Donostia-San Sebasti\'an, Spain}
\affiliation{Advanced Polymers and Materials: Physics, Chemistry and Technology, Chemistry Faculty (UPV/EHU), Paseo M. Lardizabal 3, 20018 San Sebastian, Spain}

\begin{abstract}
	We study the quantum Hall effect in a cubic lattice subjected to parallel electric and magnetic fields aligned along a crystal axis. The dual fields confine electrons in three dimensions with their classical orbits residing on the surface of a finite-size cylinder. In the quantum limit, the spectrum yields a three-dimensional generalization of the Hofstadter butterfly, featuring equidistant resonances near the spectral center and discrete levels near the boundaries. When the Fermi energy lies within the gaps between these levels, the Hall conductance in the plane normal to the fields is quantized while other components of the conductance tensor vanish. Under open boundary conditions, this quantum Hall state exhibits topological chiral hinge modes protected by bulk Chern numbers. Our results offer a novel platform for studying the quantum Hall effect in both solid-state heterostructures and synthetic lattices.
\end{abstract}
\date{\today}
\maketitle

In a three-dimensional (3D) lattice, a magnetic field parallel to the $z$-axis confines the in-plane ($xy$) electronic motion into cyclotron orbits characterized by the frequency $\omega_B$, resulting in a set of dispersive 1D Landau bands~\cite{landau1930diamagnetismus}. On the other hand, an electric field parallel to this same axis, acting together with the periodic lattice potential, causes electrons to undergo Bloch oscillations characterized by the Bloch frequency $\omega_E$~\cite{bloch1929quantenmechanik,zener1934theory}. These oscillations quantize the continuous 3D energy band into equidistant 2D bands known as the Wannier-Stark ladders~\cite{Wannier1960}. Individually, these fields only quantize the bands in one or two dimensions, leaving the spectrum dispersive along the remaining directions, hindering the formation of complete spectral gaps. However, when both electric and magnetic fields are simultaneously applied along the $z$-axis, the electronic motion is confined in all three dimensions. The associated spectrum contains discrete energy levels, which are termed Wannier-Stark-Landau (WSL) states and can be regarded as 3D flat bands~\cite{Ferreira1991,Pacheco1992,wacker1998vertical,Gluck2002,margulis2003collective}. In a simple model of free electron gas subjected to parallel fields, these levels emerge as a dual-ladder structure given by~\cite{Bass1980,claro1990}
\begin{equation}
	E_{n\nu} = \nu\hbar\omega_E + \hbar\omega_B\left(n + \frac{1}{2}\right), \label{Eq: continuum}
\end{equation}
where $\nu=0,\pm1,\pm2,\dots$ indexes the Wannier-Stark ladders and $n=0,1,2,\dots$ denotes the Landau level index. The structure of the WSL levels depends on the ratio between the two frequencies $\omega_E$ and $\omega_B$. If the frequencies are commensurate, i.e., the ratio $\omega_E/\omega_B$ is a rational number, the energy levels become infinitely degenerate and equidistant. If they are incommensurate --- $\omega_E/\omega_B$ is irrational, the spectrum becomes highly complex and may exhibit chaotic features~\cite{claro1990}. Such unique characteristics make this system an intriguing platform for observing novel phenomena. For example, the commensurate interplay between the two frequencies can be described by a Diophantine equation, giving rise to a Devil's staircase in the integrated optical absorption~\cite{claro1990}. When $\omega_E$ is an integer multiple of $\omega_B$, the two ladders align, opening new elastic transport channels that manifest as sharp peaks known as Stark-cyclotron resonances in the current-voltage characteristics~\cite{Lyanda-Geller1995,Canali1996,Shon1996,liu1997sequential,Helm1999,Bryksin1999,Blaser2000,Hyart2009,Hwang2021}. Furthermore, if the magnetic field is tilted relative to the electric field, the electron dynamics can be described as a harmonic oscillator driven by a traveling wave, whose phase plane is threaded by a stochastic web~\cite{Fromhold2001,Soskin2015}. However, the topological classification of the resulting flat bands in this system, and the transport properties this implies, have remained unexplored.
\begin{figure}
	\includegraphics[width=\linewidth]{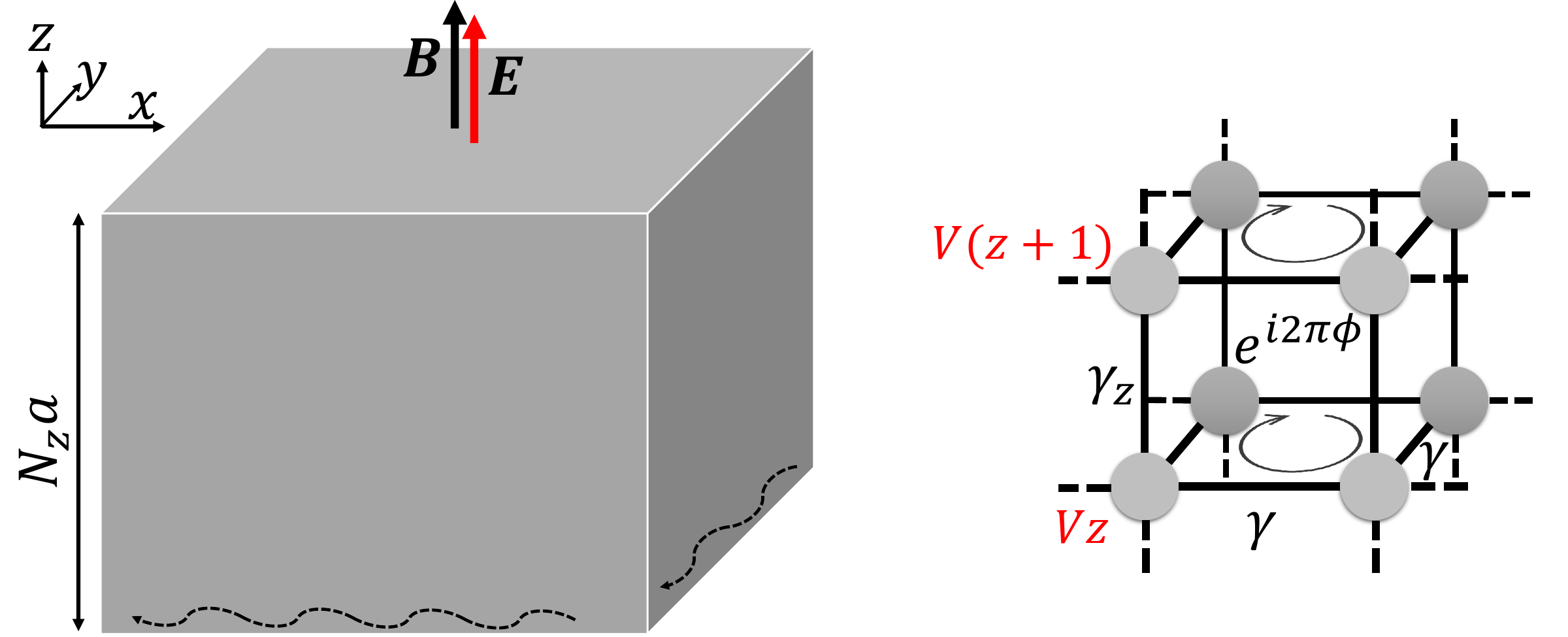}
	\caption{\label{fig:model} \textit{Cubic lattice in parallel magnetic and electric fields}. The cubic lattice is subjected to parallel magnetic and electric fields along the $z$-axis. The motion of electrons in this lattice is described by the hopping parameters $\gamma$ and $\gamma_z$, the Peierls phase, and the on-site energy $Vz$.}
\end{figure}

Here, we demonstrate that a simple 3D cubic lattice subjected to parallel fields provides a novel platform for realizing a quantum Hall effect induced by WSL states. By tracking the real-space dynamics of an electron wave packet, we map the classical orbits associated with this quantum Hall state. We compute the exact energy spectrum of the system to demonstrate a 3D generalization of the Hofstadter butterfly, featuring localized, discrete flat bands near the spectral boundaries. Importantly, when the Fermi energy resides within these spectral gaps, the Hall conductance $G_{xy}$ becomes quantized and strictly independent of the sample thickness, accompanied by the emergence of topological chiral hinge modes. Finally, we contrast our findings with other variations of the quantum Hall effect in 3D lattices.

\textit{Model.} --- We consider a cubic lattice subjected to a magnetic field $\bm{B}$ and an electric field $\bm{E}$, both aligned parallel to the $z$-axis (see Fig.~\ref{fig:model}). Using the Landau gauge, the magnetic field is described by the vector potential $\bm{A}(x,y)=(0, Bx, 0)$, while the electric field is incorporated in the potential variation per lattice constant, $V=eEa$, where $a$ denotes the lattice constant. In the independent-electron approximation, the system is described by the tight-binding Hamiltonian 
\begin{align}
	\mathcal{H} &= \sum_{x,y,z}c^{\dagger}_{xyz}Vzc_{xyz} - \sum_{x,y,z}\Big[c^{\dagger}_{(x+1)yz}\gamma c_{xyz} \nonumber\\
	&+ c^{\dagger}_{x(y+1)z}\gamma e^{-i2\pi\phi x}c_{xyz} + c^{\dagger}_{xy(z+1)}\gamma_zc_{xyz} + h.c.\Big]\label{Eq: coordinate Hamiltonian}
\end{align}
with $\gamma$ and $\gamma_z$ representing the in-plane and out-of-plane hopping amplitudes, respectively, and $\phi$ being the magnetic flux per unit cell in the $xy$-plane in units of the flux quantum $h/e$. The atomic positions $(x,y,z)$ are expressed in units of the lattice constant $a$, and we choose $a=1$ for simplicity hereafter. 

In this configuration, the discrete translation symmetry is only preserved along the $y$ direction. This symmetry is restored along the $x$ direction by restricting the magnetic flux to rational values, $\phi=p/q$, where $p$ and $q$ are coprime integers. Along the $z$-axis, the lattice terminates at $z=\pm Na$, corresponding to a finite system of $N_z=2N+1$ atomic layers. Throughout this work, we let $\gamma=1$ and $\gamma_z=0.5$. For $N_z=101$, finite-size effects along the $z$ direction are negligible compared with the energy scales of interest.

\textit{Semiclassical limit.} --- The application of parallel electric and magnetic fields localizes the crystal electrons in all three dimensions. Semiclassically, each electron executes cyclotron orbits in the $xy$-plane with frequency $\omega_B = \frac{eB}{m_e}$ while simultaneously undergoing Bloch oscillations along the $z$ direction with frequency $\omega_E=\frac{eEa}{\hbar}$. In the long wavelength limit of Hamiltonian~\eqref{Eq: coordinate Hamiltonian}, these dynamic frequencies correspond to the characteristic energies $\hbar\omega_B=4\pi|\gamma|\phi$ and $\hbar\omega_E=V$, which define the energy separations between the Landau levels in a 2D square lattice and the Wannier-Stark ladders in a 1D lattice, respectively.

The interplay between cyclotron motion and Bloch oscillations can be captured in the 3D lattice~\eqref{Eq: coordinate Hamiltonian} by tracking the dynamics of an electron wave packet. We initialize a Gaussian wave packet of the form
\begin{equation}
	W_0(\bm{r}) = \frac{1}{(2\pi\sigma^2)^{\frac{3}{4}}} \exp\left[-\frac{(\bm{r}-\bm{r}_0)^2}{4\sigma^2}\right]e^{i\bm{q}\cdot\bm{r}},
\end{equation}
where $\sigma$ represents the spatial width, $\bm{r}_0$ is the initial center-of-mass (CoM) position, and $\bm{q}$ is a wave vector that provides an initial velocity to the wave packet. The time evolution of this wave packet, $W(\bm{r},t) = \exp\left(-\frac{i}{\hbar}\mathcal{H}t\right)W_0(\bm{r})$, is computed numerically with high precision by expanding the propagator in a series of Chebyshev polynomials as~\cite{kosloff1986direct,Wang1998,fehske2007computational}
\begin{align}
	\exp\left(-\frac{i}{\hbar}\mathcal{H}t\right) &= \exp\left[-\frac{i}{\hbar}\left(a_H\tilde{H}+b_H\right)t\right] \nonumber\\
	&= e^{-ib_Ht}\left[c_0 + 2\sum_{k=1}^Mc_kT_k(\tilde{H})\right].
\end{align}
Here, $a_H$ and $b_H$ are scaling factors that bound the spectrum of the rescaled Hamiltonian $\tilde{H}$ within the interval $(-1,1)$, and $T_k(\tilde{H})$ are the Chebyshev polynomials of the rescaled Hamiltonian. The expansion coefficients are given by $c_k=(-i)^kJ_k(a_Ht/\hbar)$, where $J_k(a_Ht/\hbar)$ is the Bessel function of the first kind of order $k$. The expansion is truncated at a cutoff of $M=4096$.
\begin{figure}
	\includegraphics[width=\linewidth]{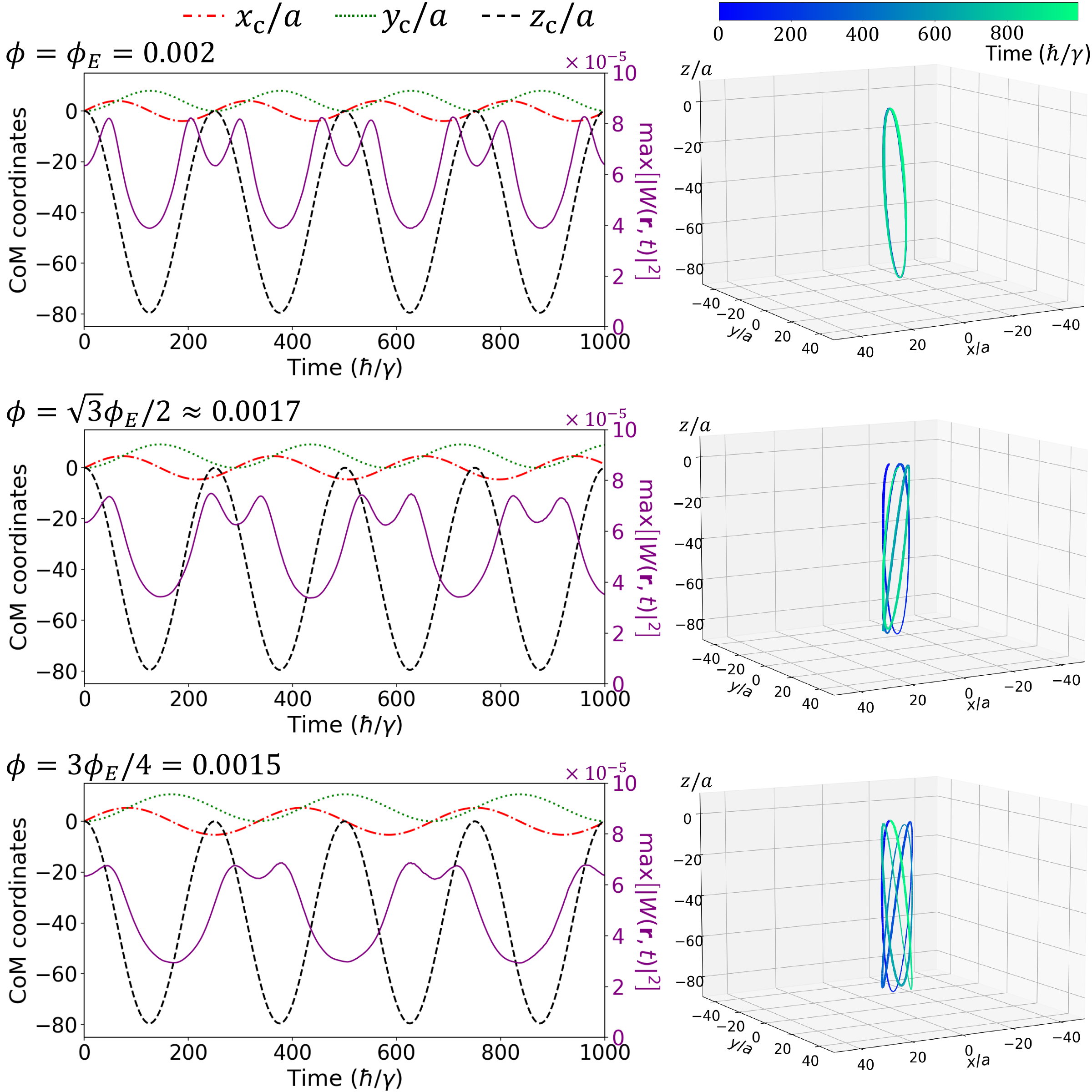}
	\caption{\label{fig:wp} \textit{Dynamics of electron wave packets}. (Left column) The CoM coordinates of the wave packet (dashed/dotted lines) and the height of its modulus squared (solid line) when $V=4\pi|\gamma|\phi_E$ with $\phi_E=0.002$. (Right column) The trajectories of the wave packet's CoM in real space. The color denotes time and the line thickness indicates the peak height of $|W(\bm{r},t)|^2$. The three rows correspond to three values of the magnetic flux $\phi$. The initial wave packet is defined by $\bm{r}_0=(0,0,0)$, $\sigma=10a$, and $\bm{q}=(0.05,0,0)/a$. The lattice dimensions are $N_x=N_y=151$, $N_z=251$.}
\end{figure}
\begin{figure*}
	\includegraphics[width=\linewidth]{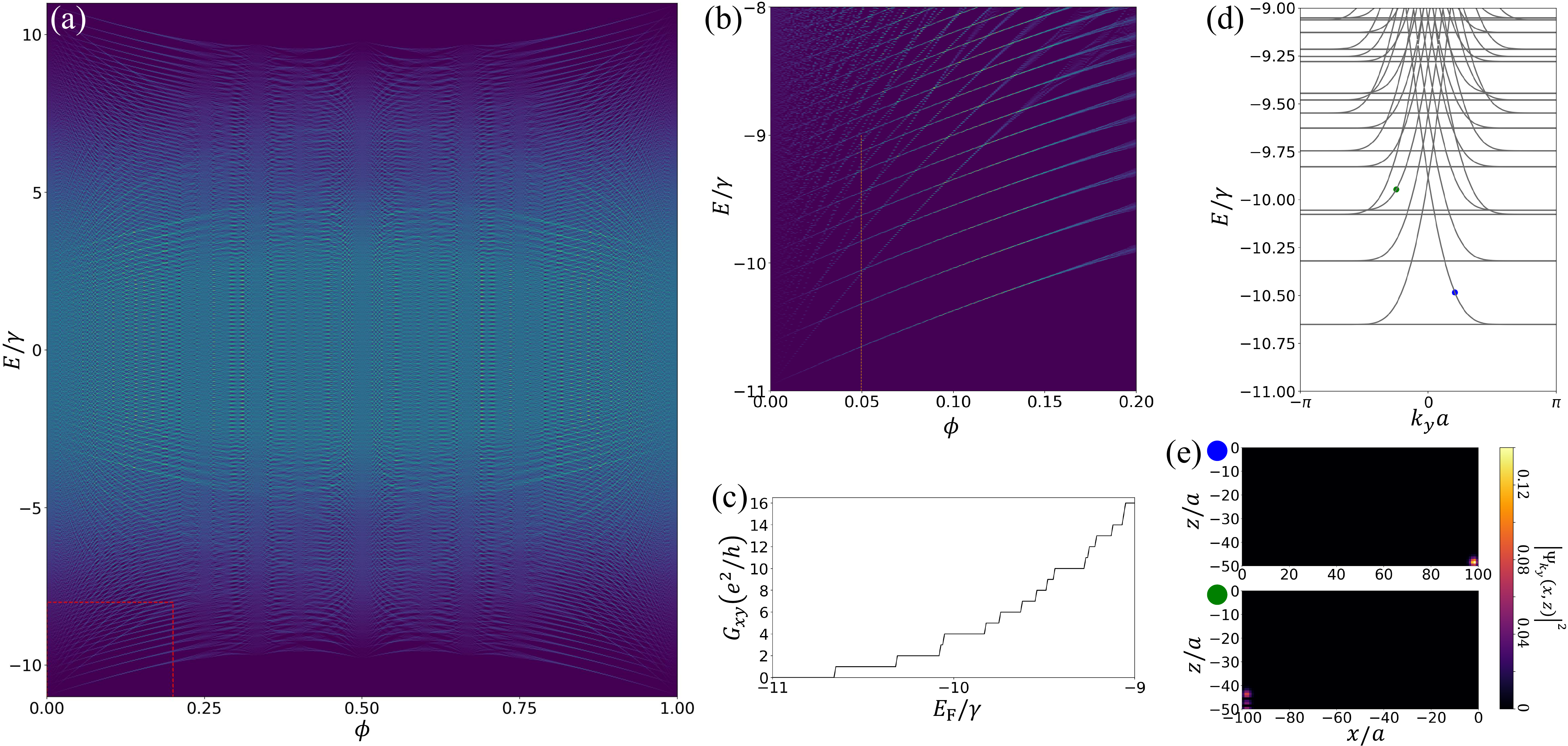}
	\caption{\label{fig:hofstadter} \textit{Three-dimensional Hofstadter-Stark spectrum and chiral hinge mode}. (a) The DOS of Hamiltonian~\eqref{Eq: momentum Hamiltonian} as a function of energy and magnetic flux when $N_z=101$ and $V=4\pi|\gamma|\times0.01$. The magnetic flux is given by a rational number, $p/q$, with $q=251$. (b) The magnified spectrum indicated by the red dashed box in (a), obtained by increasing $q$ to $997$. (c) The Hall conductance $G_{xy}$ as a function of the Fermi energy along the orange dashed line of panel (b) for $\phi=\frac{5}{101}$. (d) The dispersion of the lowest bands when the periodic boundary condition is broken along the $x$ direction with $\phi=\frac{5}{101}$ and $N_x=201$. (e) The modulus squared of two edge states' eigenmodes depicted by the blue and green dots in (d).}
\end{figure*}

The CoM position of the wave packet, $\bm{r}_{\text{c}}(t)=\braket{W(t)|\bm{r}|W(t)}$, and the maximum value of $|W(\bm{r},t)|^2$ are plotted in Fig.~\ref{fig:wp} for different values of the magnetic flux $\phi$ at a fixed potential shift $V=4\pi|\gamma|\times0.002$. To analyze the dynamics of the wave packet, it is convenient to introduce the resonant flux $\phi_E=\frac{V}{4\pi|\gamma|}=0.002$. At this resonance ($\phi=\phi_E$), the cyclotron frequency matches the Bloch frequency, aligning the two energy ladders, i.e., $\omega_B=\omega_E$. In all three cases, the CoM of the wave packet circulates in the $xy$-plane and oscillates periodically along the $z$-axis. Meanwhile, the peak of $|W(\bm{r},t)|^2$ exhibits near-periodic oscillations with the period of the in-plane cyclotron orbits, indicating periodic wave packet revivals~\cite{robinett2004quantum}. Although both cyclotron motion and Bloch oscillations lead to the wave packet revivals, the contribution from the latter is negligible compared to that of the former. 

When the magnetic flux is set to $\phi=\frac{\sqrt{3}}{2}\phi_E$, incommensurate with the resonant flux $\phi_E$, the wave packet traces an open trajectory because the timescale required for the wave packet to return to its initial position is infinite. When the magnetic flux is commensurate with $\phi_E$, the trajectories become closed cylindrical Lissajous curves. Therefore, in the absence of structural boundaries of the lattice, an electron wave packet remains confined to the surface of a finite cylinder whose geometric dimensions depend on the initial position and velocity. These bulk orbits transition into open, skipping-like trajectories if the wave packet is initialized near the lattice boundaries, where it bounces off the hard walls, in a trajectory analogous to the skipping orbits of the 2D quantum Hall effect.

\textit{Quantum limit.} --- As the discrete WSL spectrum~\eqref{Eq: continuum} was derived from the continuum and thermodynamic limits, it assumes an unbounded energy spectrum. To evaluate this spectrum under realistic lattice conditions, we examine our tight-binding model for a large but finite value of $N_z$, which naturally introduces spectral bounds. Applying the periodic boundary conditions along the $x$ and $y$ directions to eliminate the edge effects, we transform Hamiltonian~\eqref{Eq: coordinate Hamiltonian} into
\begin{align}
	\mathcal{H} &= \sum_{r,k_x,k_y,z}c^{\dagger}_{rk_xk_yz}\Big[Vz - 2\gamma\cos(2\pi\phi r+k_y)\Big]c_{rk_xk_yz} \nonumber\\
	& - \sum_{r,k_x,k_y,z}\Big[c^{\dagger}_{(r+1)k_xk_yz}\gamma c_{rk_xk_yz} + c^{\dagger}_{rk_xk_y(z+1)}\gamma_zc_{rk_xk_yz} \nonumber\\
	&+ h.c.\Big] - \sum_{k_x,k_y,z}\Big[c^{\dagger}_{1k_xk_yz}\gamma e^{iqk_x}c_{qk_xk_yz} + h.c.\Big], \label{Eq: momentum Hamiltonian}
\end{align}
where $r=1\dots q$ labels the atomic sites within the magnetic unit cell. Instead of expressing this Hamiltonian as a $qN_z\times qN_z$ matrix, we decompose it into a tensor sum of a $q\times q$ matrix and another $N_z\times N_z$ one, which facilitates the calculation of the complete spectrum and eigenstates (see Supplemental Materials). The resulting density of states (DOS) is presented in Fig.~\ref{fig:hofstadter}(a) as a function of energy and magnetic flux for a fixed potential shift $V=4\pi|\gamma|\times0.01$. Near the spectral center, $E=0$, the spectrum is nearly gapless and features equidistant resonances that manifest as discrete sharp peaks in the DOS at specific values of $\phi$. These bulk resonances become separated by spectral gaps when the electric field is strong. Meanwhile, near the spectral boundaries of Fig.~\ref{fig:hofstadter}(a), distinct gaps open between discrete energy levels, indicating the formation of flat bands within the 3D lattice. Hereafter, we focus on the 3D flat bands near the bottom spectral boundary.

As can be seen in Fig.~\ref{fig:hofstadter}(b), these flat bands can be classified into distinct families based on their evolution with the magnetic field. At low energies, certain families consist of parallel levels whose energy spacings closely resemble the Wannier-Stark ladders in a 1D lattice. These levels are parallel as they evolve linearly with respect to the magnetic field with the same slope. At slightly higher energies, other families comprise levels that converge to a single degenerate energy in the zero-flux limit, similar to the conventional 2D Landau levels. Consider the family of levels parallel to the lowest band in Fig.~\ref{fig:hofstadter}(b). Since these levels are the Wannier-Stark ladders of the same Landau level, they must be associated with the lowest Landau level $n=0$. Thus, they can be represented by $E_{0\nu} = \mathcal{E}_{\nu} + \frac{\hbar\omega_B}{2}=\mathcal{E}_{\nu} + 2\pi|\gamma|\phi$ with $\mathcal{E}_{\nu}$ being the Wannier-Stark states of a finite 1D lattice.

The formation of highly localized flat bands in this 3D lattice paves the way for a 3D quantum Hall effect. To investigate this phenomenon, we compute the Hall conductivity as a function of the Fermi energy $E_{\text{F}}$ using the Kubo-Greenwood formula~\cite{greenwood1958boltzmann,kubo1957statistical}
\begin{widetext}
	\begin{equation}
		\sigma_{\alpha\beta}(E_{\text{F}}) = -\frac{e^2}{h}\times\frac{2\pi}{\mathcal{V}} \sum_{E_m<E_{\text{F}}<E_n}\frac{2\Im\left[\braket{\Psi_m|\frac{\partial H}{\partial k_{\alpha}}|\Psi_n}\braket{\Psi_n|\frac{\partial H}{\partial k_{\beta}}|\Psi_m}\right]}{(E_m-E_n)^2},
	\end{equation}
\end{widetext}
where $\ket{\Psi_n}$ and $E_n$ are the eigenstates and eigenvalues of the Hamiltonian, $\mathcal{V}$ is the total volume of the lattice, and $\alpha,\beta\in\{x,y,z\}$. When the Fermi energy resides within a spectral gap, all the longitudinal components and most of the transverse components of the conductivity tensor vanish, except $\sigma_{xy}$. Importantly, the Hall conductance, $G_{xy} = \sigma_{xy}N_za$, becomes quantized in integer units of the conductance quantum: $G_{xy} = \frac{e^2}{h}C$, where $C$ is the Chern number evaluated over the 2D manifold $(k_x,k_y)$ --- see Fig.~\ref{fig:hofstadter}(c). These Chern numbers are in agreement with those obtained numerically by the Fukui-Hatsugai-Suzuki method~\cite{fukui2005chern}. Due to the complex crossings between different families of energy levels, the width of the Hall plateaus in Fig.~\ref{fig:hofstadter}(c) varies strongly with respect to the Fermi energy.

Notably, this quantized Hall conductance differs from both Halperin's classic condition for an intrinsic 3D quantum Hall state~\cite{halperin1987possible,Gooth2023quantum} --- which requires $\sigma_{xy}=\frac{e^2}{h}\frac{C}{d}$ for $d$ being the lattice constant along the $z$ direction --- and the recently proposed 3D quantum Hall fluid~\cite{Palumbo2026generalizedgmp}. Instead, the behavior observed here is similar to the quantum Hall effect induced by the Weyl orbits in topological semimetals~\cite{Wang2017,zhang2021cycling,Xiong2022}. To clarify this distinction, a quantized Hall conductivity typically implies that the number of chiral edge channels crossing a bulk gap scales linearly with the sample thickness, as expected for a stack of weakly interacting 2D quantum Hall layers~\cite{Bernevig2007}. In contrast, a quantized Hall conductance indicates that the number of chiral edge modes crossing a gap remains independent of the sample thickness, which can be a sign of the quantum confinement effect in the bulk or the existence of chiral hinge states~\cite{Szumniak2020,Minh2021,Fu2021,Zhang2025three}. The quantum Hall effect induced by Weyl orbits is similar to that induced by finite-size confinement in the sense that the spectral gaps shrink as the sample thickness increases~\cite{Minh2021}, but it can support chiral hinge modes~\cite{Li2020}. Unlike these cases, the quantum Hall effect in our system originates directly from the WSL states and the spectral gap is independent of the thickness $N_z$. In the continuum model whose spectrum is composed of 3D flat bands given by Eq.~\eqref{Eq: continuum}, this WSL-induced quantum Hall effect possesses a totally different quantization condition when $E_{\text{F}}=0$ that will be discussed later.

\textit{Chiral hinge modes.} --- Breaking the translation symmetry of the Hamiltonian along the $x$-axis by introducing two hard walls, we obtain the 1D energy dispersion as a function of $k_y$. The lowest energy levels are shown in Fig.~\ref{fig:hofstadter}(d), which clearly reveals the formation of topological gapless modes connecting the dispersionless bulk bands. A pair of these gapless modes with opposite group velocities constitutes a counterpropagating pair of chiral hinge channels, localized at two different hinges of the sample. The net number of chiral modes traversing a gap is exactly equal to the Chern number of that gap. By examining the wave functions of some eigenstates, we can see that these counterpropagating modes are localized specifically at the two hinges with $z=-N$ of the sample. The squared moduli of two representative hinge states at $k_ya=\pi/6$ are depicted in Fig.~\ref{fig:hofstadter}(e), which exhibit different localization lengths. As the energy of these hinge modes increases, their wave functions spread deeper into both the $x$ and $z$ directions. Their real-space propagation directions are schematically illustrated in Fig.~\ref{fig:model}. 

It is interesting to revisit the discrete spectrum~\eqref{Eq: continuum} of the continuum model for the WSL states. If we assume that the Wannier-Stark ladders remain equidistant even near the boundaries $z=\pm N$, and that all the flat bands below the charge neutrality point ($E=0$) carry the same Chern number $C_{n\nu}=1$, the Hall conductance at $E_{\text{F}}=0$ is therefore equal to the total number of occupied bands. For simplicity, we focus on the resonant regime $\phi=\phi_E$, where the condition for Stark-cyclotron resonances is met. For the Landau index $n=0$, there are $N$ levels below $E_{\text{F}}=0$. Similarly, for an index $n>0$, the number of WSL states with negative energy is $N-n$. In the end, the total number of occupied flat bands is given by $\sum_{n=0}^{N}(N-n)=\frac{N(N+1)}{2}$, which results in a Hall conductivity  $\sigma_{xy}(E_{\text{F}}=0)=\frac{e^2}{h}\frac{C}{N_za}=\frac{e^2}{h}\frac{N(N+1)}{2(2N+1)a}$. Taking the thermodynamic limit, we arrive at $\sigma_{xy}=\frac{e^2}{h}\frac{N_z}{8a}$ --- the Hall conductivity per unit length, $\sigma_{xy}/(N_za)$, approaches a universal value $\frac{e^2}{8ha^2}$ for all energy gaps near the charge neutrality point. However, in the cubic lattice model, this argument breaks down due to the finite bandwidth of the electronic spectrum: in strong magnetic and electric fields, the spectrum may have gaps near zero energy that have finite Chern numbers (see the SM for more details). Thus, if the Fermi energy lies within these gaps, the Hall conductance is still quantized.


\textit{Outlook.} --- Our findings establish WSL levels as a robust yet straightforward platform for engineering the quantum Hall effect in 3D lattices. Beyond the superlattice structures of quantum wells --- which are the traditional platform for observing the WSL states~\cite{Lyanda-Geller1995,Canali1996,Shon1996,liu1997sequential,Helm1999,Bryksin1999,Blaser2000,Hyart2009,Hwang2021} --- this phenomenon can be simulated in 2D artificial systems, such as optical, phononic, and photonic lattices by tailoring the Aubry-Andr\'{e}-Harper model~\cite{harper1955single,aubry1980analyticity} in one dimension and simulating the Bloch oscillations in the other. Finally, our results open up several compelling questions for future research, such as exploring the consequences of Landau-Zener~\cite{Helm1999} tunneling within the quantum Hall regime and investigating the coupling between the cyclotron motion and Bloch oscillations.

\textit{Acknowledgments.} --- The author acknowledges significant supports from Dario Bercioux and the administrative staff of Donostia International Physics Center. The large language models Gemini~3.1 Pro and Claude Sonnet~5 have been used to review the manuscript, optimize the codes, and develop Section~I of the Supplemental Materials.

\bibliography{qhe.bib}

@ARTICLE{Bass1980,
	author = {{Bass}, F.~G. and {Zorchenko}, V.~V. and {Shashora}, V.~I.},
	title = "{Stark-cyclotron resonance in semiconductors with a superlattice}",
	journal = {Soviet Journal of Experimental and Theoretical Physics Letters},
	year = 1980,
	month = mar,
	volume = {31},
	pages = {314},
	adsurl = {https://ui.adsabs.harvard.edu/abs/1980JETPL..31..314B}
}

@incollection{wacker1998vertical,
	title={Vertical transport and domain formation in multiple quantum wells},
	author={Wacker, Andreas},
	booktitle={Theory of transport properties of semiconductor nanostructures},
	pages={321--355},
	year={1998},
	doi={10.1007/978-1-4615-5807-1_10},
	publisher={Springer}
}

@article{Blaser2000,
	title = {Far-infrared emission and Stark-cyclotron resonances in a quantum-cascade structure based on photon-assisted tunneling transition},
	author = {Blaser, St\'ephane and Rochat, Michel and Beck, Mattias and Faist, J\'er\^ome and Oesterle, Ursula},
	journal = {Phys. Rev. B},
	volume = {61},
	issue = {12},
	pages = {8369--8374},
	numpages = {0},
	year = {2000},
	month = Mar,
	publisher = {American Physical Society},
	doi = {10.1103/PhysRevB.61.8369},
	url = {https://link.aps.org/doi/10.1103/PhysRevB.61.8369}
}

@article{Bryksin1999,
	title = {Cyclotron-Stark-phonon resonances in semiconductor superlattices under terahertz irradiation},
	author = {Bryksin, V. V. and Kleinert, P.},
	journal = {Phys. Rev. B},
	volume = {59},
	issue = {12},
	pages = {8152--8162},
	numpages = {0},
	year = {1999},
	month = Mar,
	publisher = {American Physical Society},
	doi = {10.1103/PhysRevB.59.8152},
	url = {https://link.aps.org/doi/10.1103/PhysRevB.59.8152}
}

@article{Canali1996,
	title = {Stark-Cyclotron Resonance in a Semiconductor Superlattice},
	author = {Canali, Luca and Lazzarino, Marco and Sorba, Lucia and Beltram, Fabio},
	journal = {Phys. Rev. Lett.},
	volume = {76},
	issue = {19},
	pages = {3618--3621},
	numpages = {0},
	year = {1996},
	month = May,
	publisher = {American Physical Society},
	doi = {10.1103/PhysRevLett.76.3618},
	url = {https://link.aps.org/doi/10.1103/PhysRevLett.76.3618}
}

@article{claro1990,
	title = {Novel electro-optical properties of a semiconductor superlattice under a magnetic field},
	author = {Claro, F. and Pacheco, M. and Barticevic, Z.},
	journal = {Phys. Rev. Lett.},
	volume = {64},
	issue = {25},
	pages = {3058--3061},
	numpages = {0},
	year = {1990},
	month = Jun,
	publisher = {American Physical Society},
	doi = {10.1103/PhysRevLett.64.3058},
	url = {https://link.aps.org/doi/10.1103/PhysRevLett.64.3058}
}

@article{Ferreira1991,
	title = {Resonances in the hopping probability between flexible quantum dots: The case of superlattices under parallel electric and magnetic fields},
	author = {Ferreira, R.},
	journal = {Phys. Rev. B},
	volume = {43},
	issue = {11},
	pages = {9336(R)--9338(R)},
	numpages = {0},
	year = {1991},
	month = Apr,
	publisher = {American Physical Society},
	doi = {10.1103/PhysRevB.43.9336},
	url = {https://link.aps.org/doi/10.1103/PhysRevB.43.9336}
}

@article{Fromhold2001,
	title = {Effects of Stochastic Webs on Chaotic Electron Transport in Semiconductor Superlattices},
	author = {Fromhold, T. M. and Krokhin, A. A. and Tench, C. R. and Bujkiewicz, S. and Wilkinson, P. B. and Sheard, F. W. and Eaves, L.},
	journal = {Phys. Rev. Lett.},
	volume = {87},
	issue = {4},
	pages = {046803},
	numpages = {4},
	year = {2001},
	month = Jul,
	publisher = {American Physical Society},
	doi = {10.1103/PhysRevLett.87.046803},
	url = {https://link.aps.org/doi/10.1103/PhysRevLett.87.046803}
}

@article{Gluck2002,
	title = {Wannier–Stark resonances in optical and semiconductor superlattices},
	journal = {Physics Reports},
	volume = {366},
	number = {3},
	pages = {103-182},
	year = {2002},
	issn = {0370-1573},
	doi = {https://doi.org/10.1016/S0370-1573(02)00142-4},
	url = {https://www.sciencedirect.com/science/article/pii/S0370157302001424},
	author = {Markus Glück and Andrey {R. Kolovsky} and Hans Jürgen Korsch}
}

@article{Helm1999,
	title = {Continuum Wannier-Stark Ladders Strongly Coupled by Zener Resonances in Semiconductor Superlattices},
	author = {Helm, M. and Hilber, W. and Strasser, G. and De Meester, R. and Peeters, F. M. and Wacker, A.},
	journal = {Phys. Rev. Lett.},
	volume = {82},
	issue = {15},
	pages = {3120--3123},
	numpages = {0},
	year = {1999},
	month = Apr,
	publisher = {American Physical Society},
	doi = {10.1103/PhysRevLett.82.3120},
	url = {https://link.aps.org/doi/10.1103/PhysRevLett.82.3120}
}

@article{Hwang2021,
	title = {Electric quantum oscillations in Weyl semimetals},
	author = {Hwang, Kyusung and Lee, Woo-Ram and Park, Kwon},
	journal = {Phys. Rev. Res.},
	volume = {3},
	issue = {3},
	pages = {033132},
	numpages = {12},
	year = {2021},
	month = Aug,
	publisher = {American Physical Society},
	doi = {10.1103/PhysRevResearch.3.033132},
	url = {https://link.aps.org/doi/10.1103/PhysRevResearch.3.033132}
}

@article{Lyanda-Geller1995,
	title = {Antiresonant hopping conductance and negative magnetoresistance in quantum-box superlattices},
	author = {Lyanda-Geller, Yuli and Leburton, Jean-Pierre},
	journal = {Phys. Rev. B},
	volume = {52},
	issue = {4},
	pages = {2779--2783},
	numpages = {0},
	year = {1995},
	month = Jul,
	publisher = {American Physical Society},
	doi = {10.1103/PhysRevB.52.2779},
	url = {https://link.aps.org/doi/10.1103/PhysRevB.52.2779}
}

@article{margulis2003collective,
	title={Collective electronic excitations in a semiconductor superlattice in the Landau and Wannier-Stark ladder regime},
	author={Margulis, Vl A and Makarov, SV and Piterimova, TV and Gaiduk, EA},
	journal={The European Physical Journal B-Condensed Matter and Complex Systems},
	volume={33},
	number={2},
	pages={153--164},
	year={2003},
	doi={10.1140/epjb/e2003-00152-1},
	publisher={Springer}
}

@article{Pacheco1992,
	title = {Optical response of a superlattice in parallel magnetic and electric fields},
	author = {Pacheco, M. and Barticevic, Z. and Claro, F.},
	journal = {Phys. Rev. B},
	volume = {46},
	issue = {23},
	pages = {15200--15206},
	numpages = {0},
	year = {1992},
	month = Dec,
	publisher = {American Physical Society},
	doi = {10.1103/PhysRevB.46.15200},
	url = {https://link.aps.org/doi/10.1103/PhysRevB.46.15200}
}

@article{Shon1996,
	title = {Hopping conduction in semiconductor superlattices in a quantized magnetic field},
	author = {Shon, Nguyen Hong and Nazareno, H. N.},
	journal = {Phys. Rev. B},
	volume = {53},
	issue = {12},
	pages = {7937--7944},
	numpages = {0},
	year = {1996},
	month = Mar,
	publisher = {American Physical Society},
	doi = {10.1103/PhysRevB.53.7937},
	url = {https://link.aps.org/doi/10.1103/PhysRevB.53.7937}
}

@article{Gooth2023quantum,
	title={Quantum-Hall physics and three dimensions},
	author={Gooth, Johannes and Galeski, Stanislaw and Meng, Tobias},
	journal={Reports on Progress in Physics},
	volume={86},
	number={4},
	pages={044501},
	year={2023},
	doi={10.1088/1361-6633/acb8c9},
	publisher={IOP Publishing}
}

@article{Szumniak2020,
	title = {Hinge modes and surface states in second-order topological three-dimensional quantum Hall systems induced by charge density modulation},
	author = {Szumniak, Pawe\l{} and Loss, Daniel and Klinovaja, Jelena},
	journal = {Phys. Rev. B},
	volume = {102},
	issue = {12},
	pages = {125126},
	numpages = {11},
	year = {2020},
	month = {Sep},
	publisher = {American Physical Society},
	doi = {10.1103/PhysRevB.102.125126},
	url = {https://link.aps.org/doi/10.1103/PhysRevB.102.125126}
}

@article{halperin1987possible,
	title={Possible states for a three-dimensional electron gas in a strong magnetic field},
	author={Halperin, Bertrand I},
	journal={Japanese Journal of Applied Physics},
	volume={26},
	number={S3-3},
	pages={1913},
	doi={10.7567/JJAPS.26S3.1913},
	year={1987}
}

@article{Wang2017,
	title = {3D Quantum Hall Effect of Fermi Arcs in Topological Semimetals},
	author = {Wang, C. M. and Sun, Hai-Peng and Lu, Hai-Zhou and Xie, X. C.},
	journal = {Phys. Rev. Lett.},
	volume = {119},
	issue = {13},
	pages = {136806},
	numpages = {7},
	year = {2017},
	month = {Sep},
	publisher = {American Physical Society},
	doi = {10.1103/PhysRevLett.119.136806},
	url = {https://link.aps.org/doi/10.1103/PhysRevLett.119.136806}
}

@article{Minh2021,
	title = {Quantum Hall effect induced by chiral Landau levels in topological semimetal films},
	author = {Nguyen, D.-H.-Minh and Kobayashi, Koji and Wichmann, Jan-Erik R. and Nomura, Kentaro},
	journal = {Phys. Rev. B},
	volume = {104},
	issue = {4},
	pages = {045302},
	numpages = {12},
	year = {2021},
	month = {Jul},
	publisher = {American Physical Society},
	doi = {10.1103/PhysRevB.104.045302},
	url = {https://link.aps.org/doi/10.1103/PhysRevB.104.045302}
}

@article{Li2020,
	title = {3D Quantum Hall Effect and a Global Picture of Edge States in Weyl Semimetals},
	author = {Li, Hailong and Liu, Haiwen and Jiang, Hua and Xie, X. C.},
	journal = {Phys. Rev. Lett.},
	volume = {125},
	issue = {3},
	pages = {036602},
	numpages = {6},
	year = {2020},
	month = Jul,
	publisher = {American Physical Society},
	doi = {10.1103/PhysRevLett.125.036602},
	url = {https://link.aps.org/doi/10.1103/PhysRevLett.125.036602}
}

@article{Bernevig2007,
	title = {Theory of the Three-Dimensional Quantum Hall Effect in Graphite},
	author = {Bernevig, B. Andrei and Hughes, Taylor L. and Raghu, Srinivas and Arovas, Daniel P.},
	journal = {Phys. Rev. Lett.},
	volume = {99},
	issue = {14},
	pages = {146804},
	numpages = {4},
	year = {2007},
	month = {Oct},
	publisher = {American Physical Society},
	doi = {10.1103/PhysRevLett.99.146804},
	url = {https://link.aps.org/doi/10.1103/PhysRevLett.99.146804}
}

@article{liu1997sequential,
	title={Sequential resonant tunnelling through Landau levels in GaAs/AlAs superlattices},
	author={Liu, Jian and Gornik, Erich and Xu, Shijie and Zheng, Houzhi},
	journal={Semiconductor science and technology},
	volume={12},
	number={11},
	doi={10.1088/0268-1242/12/11/015},
	pages={1422--1424},
	year={1997}
}

@article{robinett2004quantum,
	title={Quantum wave packet revivals},
	author={Robinett, Richard Wallace},
	journal={Physics reports},
	volume={392},
	number={1-2},
	pages={1--119},
	year={2004},
	publisher={Elsevier},
	doi={10.1016/j.physrep.2003.11.002}
}

@article{fukui2005chern,
	title={Chern numbers in discretized Brillouin zone: Efficient method of computing (spin) Hall conductances},
	author={Fukui, Takahiro and Hatsugai, Yasuhiro and Suzuki, Hiroshi},
	journal={Journal of the Physical Society of Japan},
	volume={74},
	number={6},
	pages={1674--1677},
	year={2005},
	doi={10.1143/JPSJ.74.1674},
	publisher={The Physical Society of Japan}
}

@article{greenwood1958boltzmann,
	title={The Boltzmann equation in the theory of electrical conduction in metals},
	author={Greenwood, DA},
	journal={Proceedings of the Physical Society},
	volume={71},
	number={4},
	pages={585--596},
	year={1958},
	doi={10.1088/0370-1328/71/4/306}
}

@article{kubo1957statistical,
	title={Statistical-mechanical theory of irreversible processes. I. General theory and simple applications to magnetic and conduction problems},
	author={Kubo, Ryogo},
	journal={Journal of the physical society of Japan},
	volume={12},
	number={6},
	pages={570--586},
	year={1957},
	publisher={The Physical Society of Japan},
	doi={10.1143/JPSJ.12.570}
}

@article{Xiong2022,
	title = {Understanding the three-dimensional quantum Hall effect in generic multi-Weyl semimetals},
	author = {Xiong, Feng and Honerkamp, Carsten and Kennes, Dante M. and Nag, Tanay},
	journal = {Phys. Rev. B},
	volume = {106},
	issue = {4},
	pages = {045424},
	numpages = {18},
	year = {2022},
	month = Jul,
	publisher = {American Physical Society},
	doi = {10.1103/PhysRevB.106.045424},
	url = {https://link.aps.org/doi/10.1103/PhysRevB.106.045424}
}

@article{zhang2021cycling,
	title={Cycling Fermi arc electrons with Weyl orbits},
	author={Zhang, Cheng and Zhang, Yi and Lu, Hai-Zhou and Xie, XC and Xiu, Faxian},
	journal={Nature Reviews Physics},
	volume={3},
	number={9},
	pages={660--670},
	year={2021},
	publisher={Nature Publishing Group UK London},
	doi={10.1038/s42254-021-00344-z}
}

@article{harper1955single,
	title={Single band motion of conduction electrons in a uniform magnetic field},
	author={Harper, Philip George},
	journal={Proceedings of the Physical Society. Section A},
	volume={68},
	number={10},
	pages={874--878},
	doi={10.1088/0370-1298/68/10/304},
	year={1955}
}

@article{aubry1980analyticity,
	title={Analyticity breaking and Anderson localization in incommensurate lattices},
	author={Aubry, Serge and Andr{\'e}, Gilles},
	journal={Ann. Israel Phys. Soc},
	volume={3},
	number={133},
	pages={18},
	year={1980}
}

@article{landau1930diamagnetismus,
	title={Diamagnetismus der metalle},
	author={Landau, LD},
	journal={Zeitschrift f{\"u}r Physik},
	volume={64},
	number={9},
	pages={629--637},
	year={1930},
	doi={10.1007/BF01397213},
	publisher={Springer}
}

@article{bloch1929quantenmechanik,
	title={{\"U}ber die quantenmechanik der elektronen in kristallgittern},
	author={Bloch, Felix},
	journal={Zeitschrift f{\"u}r physik},
	volume={52},
	number={7},
	pages={555--600},
	year={1929},
	doi={10.1007/BF01339455},
	publisher={Springer}
}

@article{zener1934theory,
	title={A theory of the electrical breakdown of solid dielectrics},
	author={Zener, Clarence},
	journal={Proceedings of the Royal Society of London. Series A, Containing Papers of a Mathematical and Physical Character},
	volume={145},
	number={855},
	pages={523--529},
	year={1934},
	publisher={The Royal Society London}
}

@article{Wannier1960,
	title = {Wave Functions and Effective Hamiltonian for Bloch Electrons in an Electric Field},
	author = {Wannier, Gregory H.},
	journal = {Phys. Rev.},
	volume = {117},
	issue = {2},
	pages = {432--439},
	numpages = {0},
	year = {1960},
	month = Jan,
	publisher = {American Physical Society},
	doi = {10.1103/PhysRev.117.432},
	url = {https://link.aps.org/doi/10.1103/PhysRev.117.432}
}

@article{Fu2021,
	title = {Bulk-hinge correspondence and three-dimensional quantum anomalous Hall effect in second-order topological insulators},
	author = {Fu, Bo and Hu, Zi-Ang and Shen, Shun-Qing},
	journal = {Phys. Rev. Res.},
	volume = {3},
	issue = {3},
	pages = {033177},
	numpages = {14},
	year = {2021},
	month = {Aug},
	publisher = {American Physical Society},
	doi = {10.1103/PhysRevResearch.3.033177},
	url = {https://link.aps.org/doi/10.1103/PhysRevResearch.3.033177}
}

@book{fehske2007computational,
	title={Computational many-particle physics},
	author={Fehske, Holger and Schneider, Ralf and Weisse, Alexander},
	year={2007},
	doi={10.1007/978-3-540-74686-7_19},
	publisher={Springer}
}

@article{kosloff1986direct,
	title={A direct relaxation method for calculating eigenfunctions and eigenvalues of the Schr{\"o}dinger equation on a grid},
	author={Kosloff, Ronnie and Tal-Ezer, H},
	journal={Chemical Physics Letters},
	volume={127},
	number={3},
	pages={223--230},
	year={1986},
	doi={10.1016/0009-2614(86)80262-7},
	publisher={Elsevier}
}

@article{Wang1998,
	title = {Time-dependent approach to scattering by Chebyshev-polynomial expansion and the fast-Fourier-transform algorithm},
	author = {Wang, J. B. and Scholz, T. T.},
	journal = {Phys. Rev. A},
	volume = {57},
	issue = {5},
	pages = {3554--3559},
	numpages = {0},
	year = {1998},
	month = {May},
	publisher = {American Physical Society},
	doi = {10.1103/PhysRevA.57.3554},
	url = {https://link.aps.org/doi/10.1103/PhysRevA.57.3554}
}

@article{Xiao2010,
	title = {Berry phase effects on electronic properties},
	author = {Xiao, Di and Chang, Ming-Che and Niu, Qian},
	journal = {Rev. Mod. Phys.},
	volume = {82},
	issue = {3},
	pages = {1959--2007},
	numpages = {0},
	year = {2010},
	month = {Jul},
	publisher = {American Physical Society},
	doi = {10.1103/RevModPhys.82.1959},
	url = {https://link.aps.org/doi/10.1103/RevModPhys.82.1959}
}

@article{Hyart2009,
	title = {Model of the Influence of an External Magnetic Field on the Gain of Terahertz Radiation from Semiconductor Superlattices},
	author = {Hyart, Timo and Mattas, Jussi and Alekseev, Kirill N.},
	journal = {Phys. Rev. Lett.},
	volume = {103},
	issue = {11},
	pages = {117401},
	numpages = {4},
	year = {2009},
	month = {Sep},
	publisher = {American Physical Society},
	doi = {10.1103/PhysRevLett.103.117401},
	url = {https://link.aps.org/doi/10.1103/PhysRevLett.103.117401}
}

@article{Soskin2015,
	title = {Regular Rather than Chaotic Origin of the Resonant Transport in Superlattices},
	author = {Soskin, S. M. and Khovanov, I. A. and McClintock, P. V. E.},
	journal = {Phys. Rev. Lett.},
	volume = {114},
	issue = {16},
	pages = {166802},
	numpages = {6},
	year = {2015},
	month = {Apr},
	publisher = {American Physical Society},
	doi = {10.1103/PhysRevLett.114.166802},
	url = {https://link.aps.org/doi/10.1103/PhysRevLett.114.166802}
}

@article{Zhang2025three,
	title={Three-dimensional quantum anomalous Hall effect in Weyl semimetals},
	author={Zhang, Zhi-Qiang and Li, Yu-Hang and Lu, Ming and Liu, Hongfang and Li, Hailong and Jiang, Hua and others},
	journal={Science Bulletin},
	volume={70},
	number={22},
	pages={3729--3732},
	year={2025},
	doi={10.1016/j.scib.2025.09.037},
	publisher={Science China Press}
}

@misc{Palumbo2026generalizedgmp,
	title={Generalized GMP Algebra for Three-Dimensional Quantum Hall Fluids of Extended Objects}, 
	author={Giandomenico Palumbo},
	year={2026},
	eprint={2602.15664},
	archivePrefix={arXiv},
	primaryClass={hep-th},
	url={https://arxiv.org/abs/2602.15664}, 
}

@article{Maharaj2017,
	title = {Hall number across a van Hove singularity},
	author = {Maharaj, Akash V. and Esterlis, Ilya and Zhang, Yi and Ramshaw, B. J. and Kivelson, S. A.},
	journal = {Phys. Rev. B},
	volume = {96},
	issue = {4},
	pages = {045132},
	numpages = {16},
	year = {2017},
	month = {Jul},
	publisher = {American Physical Society},
	doi = {10.1103/PhysRevB.96.045132},
	url = {https://link.aps.org/doi/10.1103/PhysRevB.96.045132}
}

\pagebreak

\onecolumngrid
\pagebreak
\renewcommand{\theequation}{S\arabic{equation}}
\renewcommand{\thefigure}{S\arabic{figure}}
\setcounter{equation}{0}
\setcounter{figure}{0}
\setcounter{page}{1}

\begin{center}
	\Large{\textbf{SUPPLEMENTAL MATERIALS:\\ Quantum Hall effect in three-dimensional lattice induced by Wannier-Stark-Landau localization}}
\end{center}

\tableofcontents

\section{Wave packet dynamics}
This section provides supplemental information on the dynamics of an electron wave packet in the semiclassical limit. The semiclassical motions are studied systematically, including the Bloch oscillations in a one-dimensional (1D) lattice, cyclotron orbits in a 2D square lattice, and cylindrical Lissajous motion in 3D cubic lattice. Notably, a generalization of the Landau levels is introduced in the square lattice, where the quantized energy of semiclassical orbits agrees excellently with the exact diagonalization of the Hamiltonian even beyond the long-wavelength and weak-field limits.

In general, the dynamics of a wave packet is described by the semiclassical equations of motion~\cite{Xiao2010}
\begin{align}
	\dot{\bm{r}} &= \frac{\partial\varepsilon_M(\bm{k})}{\hbar\partial\bm{k}} - \dot{\bm{k}}\times\bm{\Omega}(\bm{k}),\\
	\hbar\dot{\bm{k}} &= -e\bm{E} - e\dot{\bm{r}}\times\bm{B},
\end{align}
where $\varepsilon_M(\bm{k}) = \mathcal{E}(\bm{k}) - \bm{B}\times\bm{m}(\bm{k})$ with $E(\bm{k})$ the unperturbed band energy and $\bm{m}(\bm{k})=-\frac{e}{2}\braket{W|(\bm{r}-\bm{r}_c)\times\bm{j}|W}$ the self-rotation orbital magnetic moment of the wave packet. However, as the models that we consider here are trivial and have a single orbital, the Berry curvature $\bm{\Omega}(\bm{k})$ and the orbital magnetic moment are zero. Consequently, the equations of motion are simplified into
\begin{align}
	\dot{\bm{r}} = \frac{\partial\mathcal{E}(\bm{k})}{\hbar\partial\bm{k}},\qquad \hbar\dot{\bm{k}} = -e\bm{E} - e\dot{\bm{r}}\times\bm{B}.
\end{align}
We will analytically examine dynamics of the electron wave packet via these equations, and compare some of the results with those obtained from rigorous numerical simulations. For the numerical simulations, we consider an initial Gaussian wave packet $W_0(\bm{r})$ and examine its time evolution
\begin{equation}
	W(\bm{r},t) = \exp\left(-\frac{i}{\hbar}\mathcal{H}t\right)W_0(\bm{r}).
\end{equation}
The position of the electron wave packet is represented by its center-of-mass (CoM) position
\begin{equation}
	\bm{r}_{\text{c}} = \braket{W(\bm{r},t)|\bm{r}|W(\bm{r},t)},
\end{equation}
while its spread is characterized by a vector $\bm{\chi}=(\chi_x,\chi_y,\chi_z)$ with
\begin{equation}
	\chi_{\mu} = \sqrt{\braket{W(\bm{r},t)|{\mu}^2|W(\bm{r},t)} - \braket{W(\bm{r},t)|\mu|W(\bm{r},t)}^2},\qquad\mu\in(x,y,z).
\end{equation}
The mean energy of the wave packet is given by $\braket{W(\bm{r},t)|\mathcal{H}|W(\bm{r},t)}$.
\subsection{Bloch oscillations and Wannier-Stark ladders}
\begin{figure}
	\includegraphics[width=\linewidth]{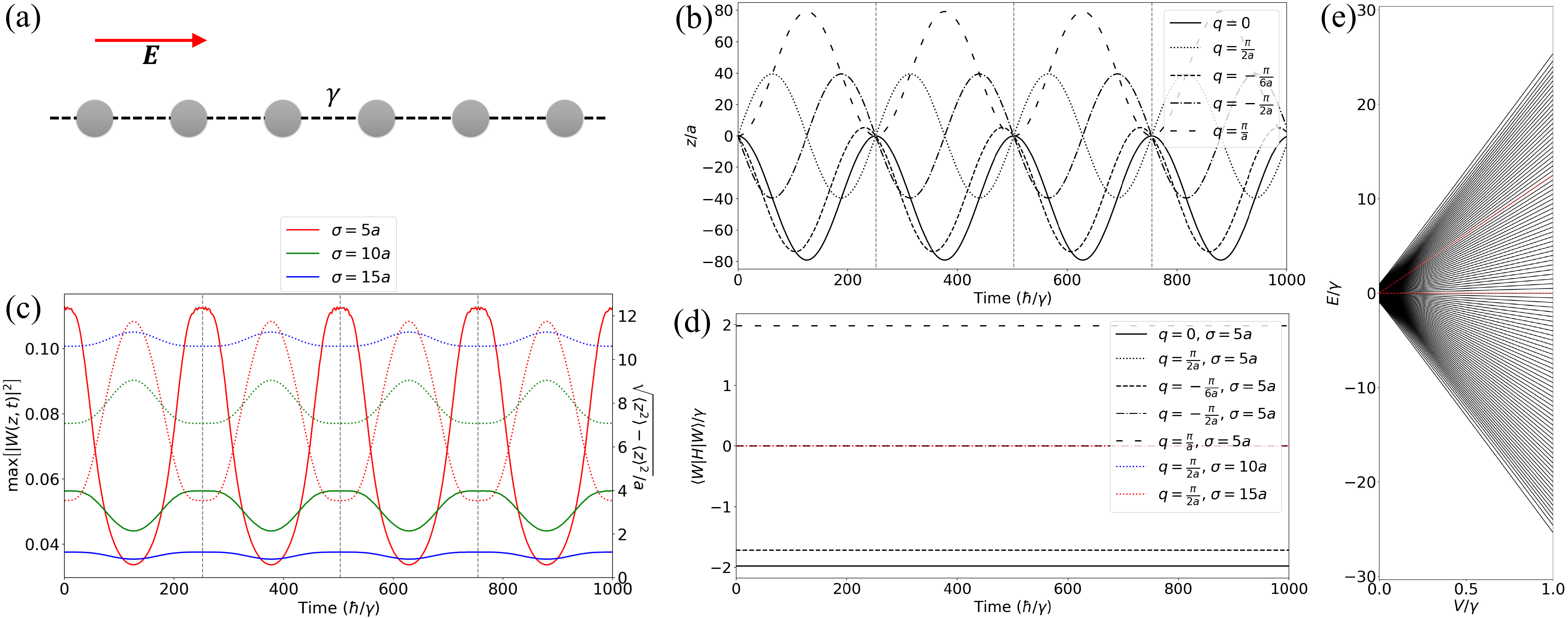}
	\caption{\label{figSM:bloch} \textit{Bloch oscillations and Wannier-Stark ladders}. (a) Sketch of the tight-binding model of a 1D lattice subjected to a parallel electric field $\bm{E}$. (b) Time dependence of the CoM position of the wave packet for different values of the initial wave vector $q$. The other parameters are $V=0.025$, $\sigma=10a$, $z_0=0$, and $N_z=501$. (c) Maximum value of $|W(z,t)|^2$ (solid lines) and spread of the wave packet (dotted lines) for different values of $\sigma$. The other parameters are $V=0.025$, $q=\frac{\pi}{2a}$, $z_0=0$, and $N_z=501$. In (b) and (c), the vertical dashed lines indicate multiples of the Bloch period $T_E=\frac{h}{V}$. (d) Mean energy of the wave packet for different cases of $q$ and $\sigma$ when $V=0.025$, $z_0=0$, and $N_z=501$. (e) Energy eigenvalues of the tight-binding model against the electric field for $N_z=101$. The two red dashed lines correspond to $\mathcal{E}=0$ and $\mathcal{E}=25V$.}
\end{figure}

Consider electrons in a 1D lattice subjected to a parallel electric field $\bm{E}$~[Fig.~\ref{figSM:bloch}(a)]. The motion of electrons is described by a tight-binding Hamiltonian with the hopping amplitude $\gamma$. If the electric field is treated as a perturbation to the system, the semiclassical equations read~\cite{Xiao2010}
\begin{equation}
	\dot{z} = \frac{\partial\mathcal{E}(k)}{\hbar\partial k},\qquad\hbar\dot{k} = -e|\bm{E}|.
\end{equation}
Here, the unperturbed dispersion is given by $\mathcal{E}(k)=-2\gamma\cos(ka)$ and $z=|\bm{r}|$. We have $k(t) = k(0) - \frac{e|\bm{E}|}{\hbar}t$ and $\dot{z} = \frac{2\gamma a}{\hbar}\sin(ka)$. Due to the periodicity of the lattice, the momentum $k$ is periodic and resides within the first Brillouin zone. Hence, we can define the Bloch period $T_E$ by identifying $k(T)-k(0)=\pm\frac{2\pi}{a}$, which yields $T_E = \frac{2\pi\hbar}{e|\bm{E}|a}$. From the group velocity given by
\begin{equation}
	v(t) = \frac{2\gamma a}{\hbar}\sin\left(-\frac{e|\bm{E}|a}{\hbar}t + k(0)a\right),
\end{equation}
we obtain the center-of-mass (CoM) position of the wave packet
\begin{equation}
	z(t) = \int dt\dot{z}(t) = \frac{2\gamma a}{e|\bm{E}|a}\cos\left(-\frac{e|\bm{E}|a}{\hbar}t + k(0)a\right) + z(0).
\end{equation}
The amplitude of oscillation depends only on the ratio of the band width and the electric field, independent of the lattice constant and the initial position. If we define the potential variation per lattice constant as $V=e|\bm{E}|a$, the Bloch period and the oscillation amplitude are expressed as $T_E=\frac{2\pi\hbar}{V}$ and $\mathcal{Z}=\frac{2\gamma a}{V}$, respectively.

This semiclassical result can be numerically illustrated by considering the dynamics of an initial Gaussian wave packet
\begin{equation}
	W_0(z) = \frac{1}{\sqrt{2\pi\sigma^2}}\exp\left[-\frac{(z-z_0)^2}{2\sigma^2}\right]e^{iqz},
\end{equation}
where $\sigma$ is the standard deviation of the wave packet and the wave vector $q$ provides it an initial velocity. We choose $\gamma=0.5$ and the lattice constant $a$ is set to unity. Tracing the time evolution of the wave packet, we obtain its CoM position, the maximum value of $|W(z,t)|^2$, its spread, and mean energy as functions shown in Fig.~\ref{figSM:bloch}(b,c,d). First, Fig.~\ref{figSM:bloch}(b) presents the CoM position for various values of initial wave vector $q$ when $V=0.025$. Despite being different in the initial conditions, all the CoM positions are sinusoidal functions with the same amplitude $\mathcal{Z}=\frac{2\times0.5a}{0.025}=40a$ and period $\frac{2\pi\hbar}{V}$, in excellent agreement with the semiclassical equations. This agreement persists even for stronger electric fields, which is beyond the semiclassical limit. Interestingly, we note that for $q=0$ and $q=\frac{\pi}{a}$, the initial velocity is identically zero but the initial accelerations are opposite in sign, leading to the difference between the two cases. Next, we consider the height and spread of the wave packets presented in Fig.~\ref{figSM:bloch}(c). In three different cases of the initial width $\sigma$, the height and spread vary periodically in time with the same period as the Bloch oscillations, suggesting periodic revivals of the wave packets. The variation is more pronounced when the initial width is smaller, which remains correct even when we plot $\frac{\max\left[|W(z,t)|^2\right]}{\max\left[|W_0(z)|^2\right]}$. Finally, the mean energy of the wave packet is plotted in Fig.~\ref{figSM:bloch}(d), which is nearly independent of the initial width $\sigma$ and varies strongly with the initial wave vector $q$. Here, we can clearly see the relation $\braket{W_0(z,q)|H|W_0(z,q)} = \mathcal{E}(q)$ that determines the energy of the wave packet.

The quantization of Bloch oscillations results in a discrete spectrum of equidistant levels, called the Wannier-Stark ladders, whose separation is $\frac{2\pi\hbar}{T_E}=\hbar\omega_E=V$. We demonstrate such a spectrum by considering the 1D tight-binding Hamiltonian
\begin{align}
	\mathcal{H}_{\text{WS}} &= \sum_{j=1}^{N_z}c^{\dagger}_jVzc_j - \sum_{j=1}^{N_z-1}\left(c^{\dagger}_{j+1}\gamma c_j + c^{\dagger}_j\gamma c_{j+1}\right).
\end{align}
We solve this Hamiltonian numerically when the lattice size is finite ($N_z$ lattice sites) and analytically when the thermodynamic limit is met. In the former case, the bounded spectrum is shown in Fig.~\ref{figSM:bloch}(e) as a function of the potential variation per lattice constant. In the absence of the electric field, $V=0$, the bandwidth is $4\gamma$. When the electric field is present, the bandwidth increases and discrete levels appear. Away from the spectral bounds, these levels evolve linearly against the electric field and are separated by $V$, as can be seen in the agreement of two levels with the red dashed lines. However, near the spectral bounds, this linear relation becomes incorrect due to the open boundary conditions of the lattice.

In the thermodynamic limit where the system size $N_z$ becomes infinite, the spectrum is unbounded. Applying the Fourier transform $c_j=\frac{1}{\sqrt{N_z}}\sum_{k}e^{-ikj}c_j$, we arrive at the following form of the Hamiltonian
$$\mathcal{H}_{\text{WS}} = \sum_{k}\left\{c^{\dagger}_{k}\frac{V}{i}\sum_{k'}\frac{\partial\delta_{k,k'}}{\partial k}c_{k'} - c^{\dagger}_{k}2\gamma\cos kc_{k}\right\},$$
where we have used the identity
$$\sum_l\sum_{k,k'}c^{\dagger}_{k}\frac{1}{N_z}le^{i(k-k')l}c_{k'} = \sum_l\sum_{k,k'}c^{\dagger}_{k}\frac{\partial}{i\partial k}\left[\frac{1}{N_z}e^{i(k-k')l}\right]c_{k'} = \sum_{k,k'}c^{\dagger}_{k}\frac{\partial\delta_{k,k'}}{i\partial k}c_{k'}.$$
This Hamiltonian can be diagonalized by introducing a set of ladder operators, $d_{\nu}$, so that $\mathcal{H}_{\text{WS}}=\sum_{\nu}\mathcal{E}_{\nu}d^{\dagger}_{\nu}d_{\nu}$. The unitary transformation between the original fermionic basis and these ladder states is given by $d^{\dagger}_{\nu} = \sum_{k}\Psi_{\nu}(k)c^{\dagger}_{k}$. Since the ladder operators satisfy the relation $\left[d_{\mu},d^{\dagger}_{\nu}\right]=\delta_{\mu,\nu}$, it is straightforward to show that $\left[\mathcal{H}_{\text{WS}}, d^{\dagger}_{\nu}\right] = \mathcal{E}_{\nu}d^{\dagger}_{\nu}$. Meanwhile, the commutator of the Hamiltonian and the ladder operator is given by
\begin{align}
	\left[\mathcal{H}_{\text{WS}}, d^{\dagger}_{\nu}\right] &= \sum_{k,k'}\left[c^{\dagger}_{k}\frac{V}{i}\sum_{\bar{k}}\frac{\partial\delta_{k,\bar{k}}}{\partial k}c_{\bar{k}} + c^{\dagger}_{k}\varepsilon(k)c_{k}, \Psi_{\nu}\left(k'\right)c^{\dagger}_{k'}\right]\\
	&= \sum_{k,k'}\Psi_{\nu}\left(k'\right)\left(\frac{V}{i}\sum_{\bar{k}}\frac{\partial\delta_{k,\bar{k}}}{\partial k}\left[c^{\dagger}_{k}c_{\bar{k}}, c^{\dagger}_{k'}\right] + \varepsilon(k)\left[c^{\dagger}_{k}c_{k}, c^{\dagger}_{k'}\right]\right)\\
	&= \sum_{k,k'}\Psi_{\nu}\left(k'\right)\left(\frac{V}{i}\sum_{\bar{k}}\frac{\partial\delta_{k,\bar{k}}}{\partial k}\delta_{\bar{k},k'}c^{\dagger}_{k} + \varepsilon(k)\delta_{k,k'}c^{\dagger}_{k}\right)\\
	&= \sum_{k,k'}\Psi_{\nu}\left(k'\right)\frac{V}{i}\frac{\partial\delta_{k,k'}}{\partial k}c^{\dagger}_{k} + \sum_{k'}\Psi_{\nu}\left(k'\right)\varepsilon(k')c^{\dagger}_{k'}\\
	&= \sum_{k}\frac{V}{i}\frac{\partial}{\partial k}\left[\sum_{k'}\Psi_{\nu}\left(k'\right)\delta_{k,k'}\right]c^{\dagger}_{k} + \sum_{k'}\Psi_{\nu}\left(k'\right)\varepsilon(k')c^{\dagger}_{k'}\\
	&= \sum_{k'}\left[\frac{V}{i}\frac{\partial\Psi_{\nu}\left(k'\right)}{\partial k'} + \Psi_{\nu}\left(k'\right)\varepsilon(k')\right]c^{\dagger}_{k'},
\end{align}
with $\varepsilon(k) = - 2\gamma\cos k$. Matching this result with $\mathcal{E}_{\nu}d^{\dagger}_{\nu} = \mathcal{E}_{\nu}\sum_{k'}\Psi_{\nu}\left(k'\right)c^{\dagger}_{k'}$, we obtain
$$\frac{V}{i}\frac{\partial\Psi_{\nu}\left(k\right)}{\partial k} - 2\gamma\Psi_{\nu}\left(k\right)\cos k = \mathcal{E}_{\nu}\Psi_{\nu}\left(k\right),$$
which can be rearranged as follows
$$\frac{1}{\Psi_{\nu}(k)}\frac{\partial\Psi_{\nu}\left(k\right)}{\partial k} = \frac{i}{V}\Big[\mathcal{E}_{\nu} + 2\gamma\cos k\Big].$$
The solution is given by
$$\Psi_{\nu}(k) = A\exp\left\{\frac{i}{V}\Big[\Big(\mathcal{E}_{\nu} + 2\gamma\sin k\Big]\right\}.$$
Here, $A$ is generally a complex amplitude, but we choose $A=1$ for the normalization condition. Although the translation symmetry of the lattice is broken along the $z$ direction due to the electric field, we impose the condition that $\Psi_{\nu}(k)$ is single-valued and periodic over the Brillouin zone, i.e., $\Psi_{\nu}(k+2\pi)=\Psi_{\nu}(k)$. This leads to the formation of the Wannier-Stark ladders in the dispersion
$$\frac{2\pi}{V}\mathcal{E}_{\nu} = 2\pi\nu \Leftrightarrow \boxed{\mathcal{E}_{\nu} = V\nu}\qquad\text{with }\nu\in\mathbb{Z},$$
which results in
$$\boxed{\Psi_{\nu}(k) = \exp\left\{i\left(k\nu + \frac{2\gamma}{V}\sin k\right)\right\}}.$$
The wave function can be obtained by considering the unitary transformation between the bases. We have
\begin{align}
	d^{\dagger}_{\nu} &= \sum_{k}\Psi_{\nu}(k)c^{\dagger}_{k} = \sum_{k}\Psi_{\nu}(k)\frac{1}{\sqrt{N_z}}\sum_{l}e^{ikl}c^{\dagger}_{l}\\
	&\rightarrow\sum_{l}\frac{N_z}{2\pi\sqrt{N_z}}\left[\int_{-\pi}^{+\pi}dk\Psi_{\nu}(k)e^{ikl}\right]c^{\dagger}_{l} = \frac{\sqrt{N_z}}{2\pi}\sum_{l}\left[\int_{-\pi}^{+\pi}dke^{i[(\nu+l)k + 2\gamma\sin k]}\right]c^{\dagger}_{l}
\end{align}
Using the standard integral representation of the Bessel function of the first kind of integer order $m$, i.e., $J_m(x) = \frac{1}{2\pi}\int_{-\pi}^{+\pi}d\theta e^{i(m\theta - x\sin\theta)}$, we can write the transformation as
\begin{equation}
	d^{\dagger}_{\nu} = \sqrt{N_z}\sum_{l}J_{\nu+l}\left(-\frac{2\gamma}{V}\right)c^{\dagger}_{l}.
\end{equation}
\subsection{Lattice cyclotron orbits and Landau levels}
Consider electrons in a 2D square lattice subjected to a perpendicular magnetic field $\bm{B}$. The motion of electrons is described by a tight-binding Hamiltonian with the hopping amplitude $\gamma$. If the magnetic field is sufficiently weak, it is treated as a perturbation to the system and we can examine the electron dynamics via the semiclassical equations
\begin{align}
	\dot{\bm{r}} = \frac{\partial\mathcal{E}(\bm{k})}{\hbar\partial\bm{k}},\qquad\hbar\dot{\bm{k}} = - e\dot{\bm{r}}\times\bm{B}.
\end{align}
The unperturbed dispersion is now given by $\mathcal{E}(\bm{k}) = -2\gamma\left[\cos(k_xa) + \cos(k_ya)\right]$, resulting in the group velocity $\dot{\bm{r}}=v_x\hat{x} + v_y\hat{y}$ with
\begin{equation}
	v_x = \frac{2\gamma a}{\hbar}\sin(k_xa)\quad\text{and}\quad v_y = \frac{2\gamma a}{\hbar}\sin(k_ya).
\end{equation}
Inserting these expressions into the other equation of motion, we arrive at
\begin{align}
	\hbar\dot{k}_x &= -ev_yB = -\frac{2e\gamma aB}{\hbar}\sin(k_ya),\\
	\hbar\dot{k}_y &= ev_xB = \frac{2e\gamma aB}{\hbar}\sin(k_xa).
\end{align}
To solve these equations analytically, we first simplify the notation by letting $k_xa\rightarrow k_x$, $k_ya\rightarrow k_y$, $\epsilon = -\frac{\mathcal{E}}{2\gamma}$, and $\omega_0=\frac{2e\gamma a^2B}{\hbar^2}$. The equations are then reduced to
\begin{align}
	\epsilon &= \cos k_x + \cos k_y,\\
	\dot{k}_x &= -\omega_0\sin k_y,\\
	\dot{k}_y &= \omega_0\sin k_x.
\end{align}
Next, we decouple the last two equations by rotating the coordinates by $\pi/4$ in momentum space and defining
\begin{equation}
	u = \frac{k_x+k_y}{2},\qquad v = \frac{k_x-k_y}{2},
\end{equation}
which are given by
\begin{align}
	\dot{u} &= \frac{\dot{k}_x+\dot{k}_y}{2} = \frac{\omega_0}{2}\left(\sin k_x - \sin k_y\right) = \omega_0\cos u\sin v,\label{EqSM: udot}\\
	\dot{v} &= \frac{\dot{k}_x-\dot{k}_y}{2} = -\frac{\omega_0}{2}\left(\sin k_x + \sin k_y\right) = -\omega_0\sin u\cos v.\label{EqSM: vdot}
\end{align}
In these new coordinates, the dispersion becomes $\epsilon = 2\cos u\cos v$. Substituting $\cos v = \frac{\epsilon}{2\cos u}$ into the square of Eq.~\eqref{EqSM: udot} yields
\begin{equation}
	\dot{u}^2 = \omega_0^2\cos^2u\left(1 - \cos^2v\right) = \omega_0^2\cos^2u\left(1 - \frac{\epsilon^2}{4\cos^2u}\right) = \omega_0^2\left(1 - \frac{\epsilon^2}{4} - \sin^2u\right).\label{EqSM: kinetic cyclotron}
\end{equation}
As $\epsilon$ varies within the range $[-2,2]$, we can define the elliptic modulus $\kappa=\sqrt{1-\frac{\epsilon^2}{4}}\in[0,1]$. Eq.~\eqref{EqSM: kinetic cyclotron} constrains the classical trajectory of the electron wave packet, $\sin^2u\le\kappa^2$. The points where $\sin u = \pm \kappa$ are the classical turning points. If the electron tried to move to a point in phase space where $\sin(u) > \kappa$, its kinetic energy would become negative, which is classically forbidden. Taking the square root of Eq.~\eqref{EqSM: kinetic cyclotron}, we arrive at the standard equation of a nonlinear (large-angle) pendulum
\begin{equation}
	\dot{u} = \omega_0\sqrt{\kappa^2 - \sin^2u},\qquad 0\le\kappa\le1,
\end{equation}
with the initial conditions $u(0)=u_0$ and $\dot{u}(0)=\dot{u}_0$. Rearranging this equation and integrating both sides, we obtain
\begin{equation}
	\int_{u_0}^{u(t)}\frac{du}{\sqrt{\kappa^2 - \sin^2u}} = \omega_0\int_0^t dt. 
\end{equation}
The integral on the left side can be transformed into the incomplete elliptic integral of the first kind by changing the variable as $\sin u = \kappa\sin\phi$ for $\phi\in[-\pi/2,+\pi/2]$. We have
\begin{equation}
	\int_{\phi_0}^{\phi(t)}\frac{d\phi}{\sqrt{1 - \kappa^2\sin^2\phi}} = \omega_0t \quad\Rightarrow\quad \phi(t) \quad\Rightarrow\quad \int_{0}^{\phi(t)}\frac{d\phi}{\sqrt{1 - \kappa^2\sin^2\phi}} = \omega_0t + \underbrace{\int_{0}^{\phi_0}\frac{d\phi}{\sqrt{1 - \kappa^2\sin^2\phi}}}_{\xi_0},
\end{equation}
whose solution is expressed in terms of the elliptic sine function $\text{sn}$ as
\begin{equation}
	\sin[\phi(t)] = \text{sn}(\omega_0t+\xi_0,\kappa)\quad\Rightarrow\quad\boxed{\sin[u(t)] = \kappa\text{sn}(\omega_0t+\xi_0,\kappa)}\label{EqSM: sine}.
\end{equation}
The parameter $\xi_0$ is uniquely determined via the initial conditions $u(0)=u_0$ and $\dot{u}(0)=\dot{u}_0$. From the solution~\eqref{EqSM: sine}, we have $\cos[u(t)] = \text{dn}(\omega_0t+\xi_0)$ and $\cos[v(t)] = \frac{\epsilon}{2\text{dn}(\omega_0t+\xi_0)}$, where $\text{dn}$ is the delta amplitude function. To determine $\sin(v(t))$, we substitute $\cos(u) = \text{dn}(\omega_0t+\xi_0, \kappa)$ into the velocity equation~\eqref{EqSM: udot} as follows 
$$\omega_0 \sin[v(t))]\text{dn}(\omega_0t+\xi_0, \kappa) = \frac{d}{dt} \left[ \arcsin(\kappa \text{sn}(\omega_0t+\xi_0, \kappa)) \right] = \frac{\omega_0 \kappa \, \text{cn}(\omega_0t+\xi_0, \kappa)\text{dn}(\omega_0t+\xi_0, \kappa)}{\sqrt{1 - \kappa^2\text{sn}^2(\omega_0t+\xi_0, \kappa)}} = \omega_0 \kappa \text{cn}(\omega_0t+\xi_0,\kappa)$$
and obtain
\begin{equation}
	\sin[v(t)] = \frac{\kappa \, \text{cn}(\omega_0t+\xi_0, \kappa)}{\text{dn}(\omega_0t+\xi_0, \kappa)}.
\end{equation}
From these expressions, we can find the functions $u(t)$ and $v(t)$. For $u(t)$, the variable is bounded inside a single oscillatory well, i.e., $|u| \le \arcsin\kappa \le \pi/2$ for all closed orbits, so that a standard single-valued $\arcsin$ branch is exact
\begin{equation}
	u(t) = \arcsin \left[ \kappa \text{sn}(\omega_0 t + \xi_0, \kappa) \right].
\end{equation}
For $v(t)$, the trajectory can wrap across multiple quadrants depending on whether it tracks an electron pocket, hole pocket, or open trajectory. To compute this without branch-cut jumps, we use the two-argument arctangent ($\text{atan2}(y, x)$), passing $\sin[v(t)]$ as $y$ and $\cos[v(t)]$ as $x$, which gives $v(t) = \text{atan2}\left( \frac{\kappa \text{cn}(\omega_0 t + \xi_0, \kappa)}{\text{dn}(\omega_0 t + \xi_0, \kappa)}, \frac{\varepsilon}{2 \text{dn}(\omega_0 t + \xi_0, \kappa)} \right)$. However, since $\text{dn}(\omega_0 t + \xi_0, \kappa) > 0$ for all real arguments on closed orbits ($\kappa < 1$), it acts as a strictly positive scaling factor and can be cleared from both arguments safely
\begin{equation}
	v(t) = \text{atan2}\left(\kappa \text{cn}(\omega_0 t + \xi_0, \kappa), \frac{\varepsilon}{2} \right).
\end{equation}
Then, we obtain the momenta of the wave packet
\begin{align}
	k_x(t) = \frac{1}{a}\left[ \arcsin \left[ \kappa \, \text{sn}(\omega_0 t + \xi_0, \kappa) \right] + \text{atan2}\left( \kappa \, \text{cn}(\omega_0 t + \xi_0, \kappa), \, \frac{\varepsilon}{2} \right) \right],\\
	k_y(t) = \frac{1}{a} \left[ \arcsin \left[ \kappa \, \text{sn}(\omega_0 t + \xi_0, \kappa) \right] - \text{atan2}\left( \kappa \, \text{cn}(\omega_0 t + \xi_0, \kappa), \, \frac{\varepsilon}{2} \right) \right].
\end{align}
Finally, substituting these continuous momentum solutions into the real-space positions obtained from the equations of motion,
\begin{align}
	x(t) = x(0) + l_B^2 [k_y(t) - k_y(0)],\\
	y(t) = y(0) - l_B^2 [k_x(t) - k_x(0)],
\end{align}
we arrive at the final precise time evolution equations for the CoM position
\begin{align}
	x(t) = x(0) + \frac{l_B^2}{a} \left[ \arcsin \left[ \kappa \, \text{sn}(\omega_0 t + \xi_0, \kappa) \right] - \text{atan2}\left( \kappa \, \text{cn}(\omega_0 t + \xi_0, \kappa), \, \frac{\varepsilon}{2} \right) - k_y(0) \right],\\
	y(t) = y(0) - \frac{l_B^2}{a} \left[ \arcsin \left[ \kappa \, \text{sn}(\omega_0 t + \xi_0, \kappa) \right] + \text{atan2}\left( \kappa \, \text{cn}(\omega_0 t + \xi_0, \kappa), \, \frac{\varepsilon}{2} \right) - k_x(0) \right].
\end{align}
The time period of electron motion is $\bar{T}_B = \frac{4K(\kappa)}{\omega_0}$ with $K(\kappa)$ being the complete elliptic integral of the first kind~\cite{Maharaj2017}. This period diverges when $\mathcal{E}=0$ since the Fermi surface becomes a square connecting the Brillouin zone boundaries, and the electron takes infinite time to reach the van Hove singularities ($M$ points). We rewrite the parameters in terms of the magnetic flux per unit cell $\phi=\frac{Ba^2}{h/e}$ as follows
\begin{equation}
	\omega_0 = \frac{4\pi\gamma}{\hbar}\frac{a^2B}{h/e} = \frac{4\pi\gamma\phi}{\hbar},\qquad l_B^2 = \frac{\hbar}{eB} = \frac{ha^2}{2\pi eBa^2} = \frac{a^2}{2\pi\phi},\qquad \kappa = \sqrt{1 - \frac{\mathcal{E}^2}{16\gamma^2}}.
\end{equation}
It is interesting to note that the real-space oscillation amplitude (or radius of the orbits) is directly tied to the maximum geometric boundaries of the crystal momentum contour. For any closed pocket at energy $\mathcal{E}$, the momentum boundary occurs when one of the directional components crosses zero. This simplifies the maximum momentum span to $\Delta(k_xa) = 2\arccos\left(\frac{|\mathcal{E}|}{2} - 1\right)$. Multiplying this by our real-space scaling factor $\frac{1}{2\pi\phi}$ yields the exact analytical formula for the spatial oscillation amplitudes
\begin{equation}
	R = R_x = R_y = \frac{\arccos\left(\frac{|\mathcal{E}|}{2} - 1\right)}{2\pi\phi}.
\end{equation}

\begin{figure}
	\includegraphics[width=\linewidth]{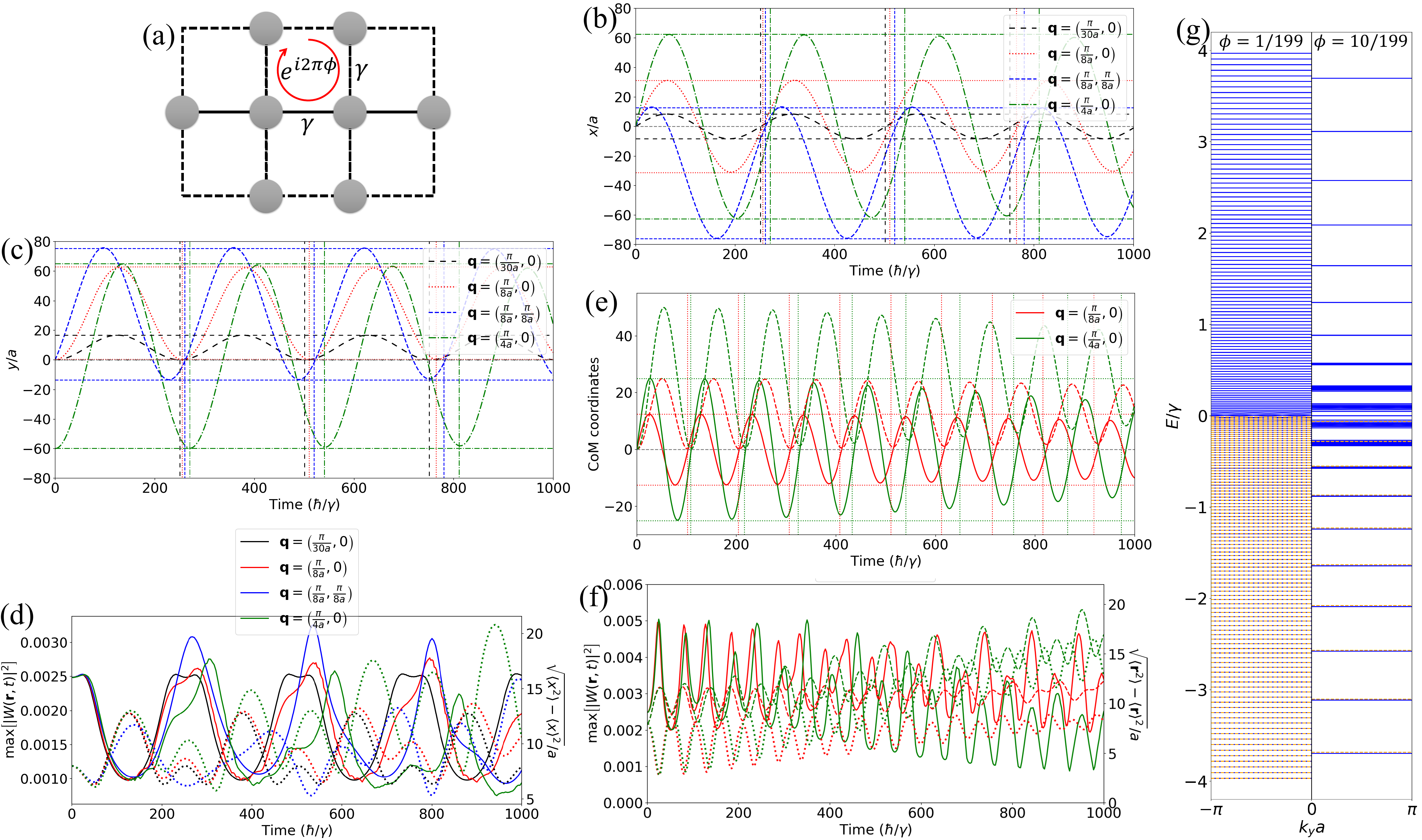}
	\caption{\label{figSM:cyclotron} \textit{Lattice cyclotron orbits and Landau levels}. (a) Sketch of the tight-binding model of a 2D square lattice subjected to a perpendicular magnetic field $\bm{B}$. The wave function of an electron that hops around a unit accumulates a phase proportional to the magnetic flux $\phi$. (b,c) Time dependence of the CoM position ($x$,$y$) of the wave packet for different values of the initial wave vector $\bm{q}$. (d) Maximum value of $|W(z,t)|^2$ (solid lines) and spread of the wave packet along the $x$ direction (dotted lines) for different values of $\bm{q}$. The other parameters for (b,c,d) are $\phi=0.002$, $\sigma=8a$, and $N_x=N_y=201$. In (b) and (c), the vertical lines indicate multiples of the semiclassical period $\bar{T}_B$, while the horizontal lines denotes the semiclassical oscillation amplitudes $R$. (e) Time dependence of the CoM position of the wave packet for two different values of $\bm{q}$ when the magnetic field is stronger. The solid and dashed lines denote the $x$ and $y$ coordinates, respectively. (f) Maximum value of $|W(z,t)|^2$ (solid lines) and spread of the wave packet along the $x$ (dotted lines) and $y$ (dashed lines) directions. The parameters for (e,f) are $\phi=0.005$, $\sigma=8a$, and $N_x=N_y=201$. (g) Spectrum of the tight-binding model at $k_x=0$ (blue solid lines) and the semiclassical quantization of energy for $E<0$ (orange dotted lines) when $\phi=1/199\approx0.005$ and $\phi=10/199\approx0.050$.}
\end{figure}

The semiclassical time period $\bar{T}_B$ and oscillation amplitude $R$ are compared with numerical simulations, which are obtained by examining the dynamics of an initial Gaussian wave packet
\begin{equation}
	W_0(\bm{r}) = \frac{1}{(2\pi\sigma^2)^{\frac{3}{4}}} \exp\left[-\frac{(\bm{r}-\bm{r}_0)^2}{4\sigma^2}\right]e^{i\bm{q}\cdot\bm{r}}.
\end{equation}
in a tight-binding model of the square lattice [Fig.~\ref{figSM:cyclotron}(a)]. The magnetic field enters the model via the Peierls substitution with a vector potential $\bm{A}(x,y)=(0,Bx,0)$ in the Landau gauge. The time evolution of the wave packet is shown in Fig.~\ref{figSM:cyclotron}. In a weak magnetic field ($\phi=0.002$) [Fig.~\ref{figSM:cyclotron}(b,c)], the wave packet motion agrees excellently with the semiclassical results in both time period and oscillation amplitude even for a large initial wave vector $\bm{q}$, which is beyond the continuum limit. However, this agreement only persists for the first few periods when $\bm{q}$ is large --- the smaller the initial wave vector is, the longer this agreement takes place. The semiclassical results diverge from the simulation when the initial wave vector $\bm{q}$ becomes the $M$ point of the Brillouin zone, i.e., the van Hove singularity. Here, the semiclassical period $\bar{T}_B$ becomes infinite but it is finite in the simulation. 

On the other hand, Fig.~\ref{figSM:cyclotron}(d) shows the height of $|W(\bm{r},t)|^2$ and the spread of the wave packet along the $x$ direction, which demonstrates the periodic wave packet revivals with the height peaks twice per period for a small value of $\bm{q}$. These periodic revivals disappear together with the two peaks per period for greater values of $\bm{q}$ and the wave packet tends to disperse more quickly. Meanwhile, in Fig.~\ref{figSM:cyclotron}(e,f), the wave packet moves in a stronger magnetic field ($\phi=0.005$), leading to shorter time periods. It is clear that the cyclotron radius decreases over time while the time periods slightly mismatch those obtained by the semiclassical equations. The peak height of the wave packet exhibits an overall declining trend, and it decays faster for $\bm{q}=\left(\frac{\pi}{4a},0\right)$ than $\bm{q}=\left(\frac{\pi}{8a},0\right)$. Interestingly, in both cases, the two-peak-per-period feature persists but one of the two peaks becomes clearly lower than the other. 

Finally, we generalize the Landau quantization by using the semiclassical period $\bar{T}_B=\frac{4K(\kappa)}{\omega_0}=\mathcal{T}_B(\mathcal{E})$ as a function of energy. Here, $\mathcal{E}$ denotes the energy at the center of a gap given by $\frac{2\pi\hbar}{\mathcal{T}_B(\mathcal{E})}$. Applying the Bohr-Sommerfeld quantization condition with the Maslov index $\nu=2$ and zero Berry phase, we obtain the energy levels
\begin{equation}
	\mathcal{E}_n = -4|\gamma| + \frac{2\pi\hbar}{\mathcal{T}_B(\mathcal{E})}\left(n + \frac{1}{2}\right).
\end{equation}
The lowest level is $\mathcal{E}_0 = -4|\gamma| + \frac{2\pi\hbar}{\mathcal{T}_B(-4|\gamma|}$ and the higher ones are then determined by
\begin{equation}
	\mathcal{E}_n = -4|\gamma| + \frac{2\pi\hbar}{\mathcal{T}_B\left(\frac{\mathcal{E}_{n}+\mathcal{E}_{n-1}}{2}\right)}\left(n + \frac{1}{2}\right).
\end{equation}
These quantized energies are shown in Fig.~\ref{figSM:cyclotron}(g) in comparison with the energy eigenvalues obtained from the tight-binding Hamiltonian at $k_x=0$. Except for levels near zero energy, where the semiclassical period diverges, the semiclassical quantization agrees excellently with the full quantum approach. The agreement extends beyond the low energy Landau levels and even beyond the weak field regime. We therefore term them the generalized Landau levels in a lattice.
\subsection{Stark-cyclotron orbits}
Consider an electron moving in a continuous space with a periodic potential $V(z)$ and parallel electric and magnetic fields along the $z$ direction. The lattice constant of the periodic potential is $a$, i.e., $V(z+a)=V(z)$. In the Landau gauge $\bm{A}(\bm{r}) = (0,Bx,0)$, the Hamiltonian of this electron reads
\begin{equation}
	\mathcal{H}_{\text{free}} = \frac{p_x^2 + (p_y + m_e\omega_Bx)^2 + p_z^2}{2m_e} + V(z) + eEz,
\end{equation}
whose eigenvalues define the Wannier-Stark-Landau (WSL) states
\begin{equation}
	E_{n\nu} = \nu\hbar\omega_E + \hbar\omega_B\left(n + \frac{1}{2}\right).
\end{equation}
Here, the Wannier-Stark ladders are indexed by $\nu=0,\pm1,\pm2,\dots$ while Landau levels are denoted by $n=0,1,2,\dots$. Within the tight-binding approximation for the periodic potential, the coefficients of the cell eigenfunctions are governed by recursion relations satisfied by Bessel functions of the first kind. The argument of these functions represents the ratio of the inter-well coupling (miniband width $2\gamma$) to the field-induced detuning ($eEa$). In the high-field regime ($eEa\gg2\gamma$), the argument becomes small, leading to a faster-than-exponential spatial localization of the WSL states. 

Stark-cyclotron resonance is achieved when the ratio between the energy separation of the Wannier-Stark ladder and the cyclotron energy is rational: $\frac{\omega_E}{\omega_B}=\frac{r}{w}$ for $r,w\in\mathbb{Z}^+$. In a semiconductor superlattice structure, the physical manifestation of these resonances is governed by a Diophantine equation, which predicts an infinite number of solutions when the reduced magnetic energy is rational. This Diophantine link is the mathematical origin of the fractal spectral structures; the integrated absorption forms a ``Devil’s Staircase" because the resonance positions are sensitive to the commensurate or incommensurate nature of the frequency ratio~\cite{claro1990}.

In a cubic lattice subjected to two parallel fields, the condition for Stark-cyclotron resonance is modified by replacing the continuum cyclotron frequency $\omega_B$ with the lattice one $\bar{\omega}_B=\frac{2\pi}{\bar{T}_B}$, i.e.,
\begin{equation}
	\frac{\omega_E}{\bar{\omega}_B}=\frac{r}{w}.
\end{equation}
However, as this condition now depends on the electron energy, the Bloch frequency $\omega_E$ can only be commensurate with the cyclotron frequency at a specific energy and becomes incommensurate elsewhere. For example, consider the lattice Landau levels when $\phi=1/199$ [Fig.~\ref{figSM:cyclotron}(g)], if $\hbar\omega_E$ is equal to the spacing between Landau levels at the spectral bounds, it is greater than the level separations near the spectral center. Therefore, in this cubic lattice, the commensurability between the two frequencies no longer yields a discrete spectrum with equally spaced levels. In fact, if the two fields are relatively weak, the spectrum is approximately gapless. The dependence of the spectrum on the electric and magnetic fields will be examined in the next section.

\begin{figure}
	\includegraphics[width=\linewidth]{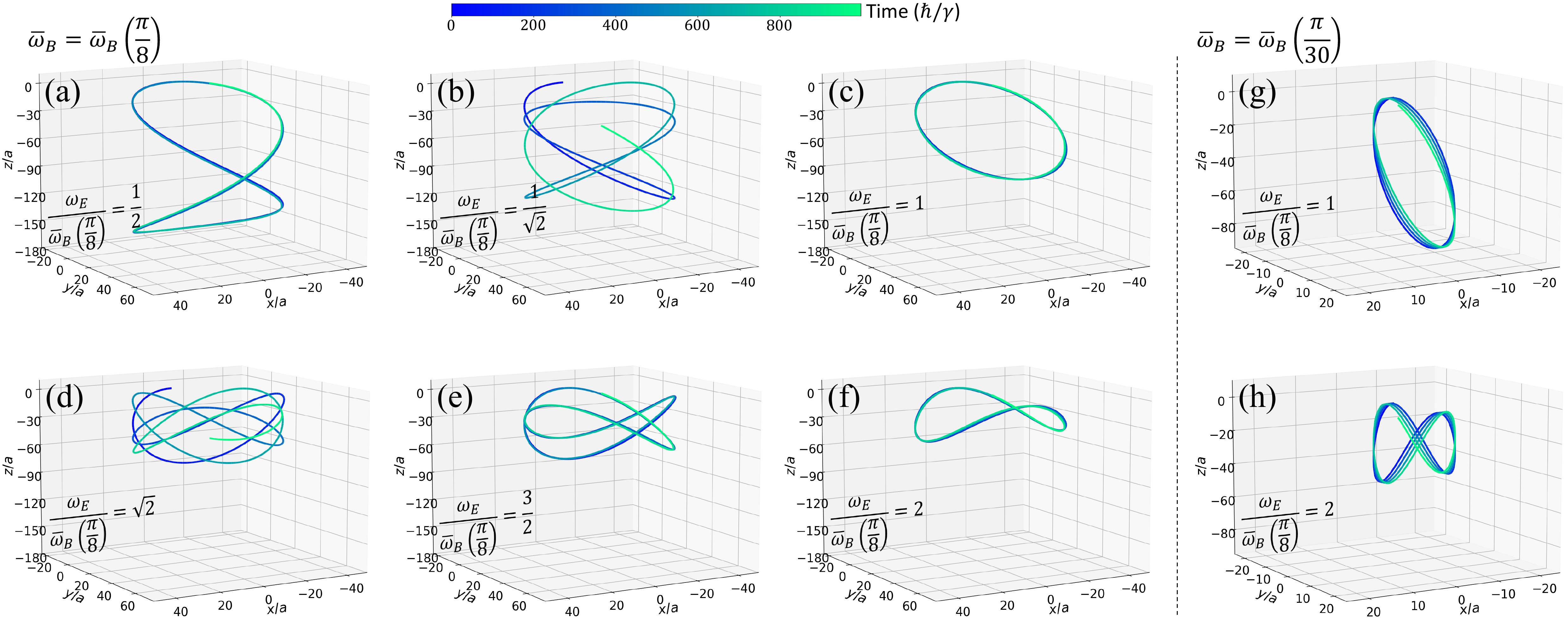}
	\caption{\label{figSM:stark-cyclotron} \textit{Semiclassical orbits of WSL states}. (a-f) Semiclassical trajectory of the wave packet for different values of electric field when the magnetic flux is $\phi=0.002$ and the initial wave vector is $\bm{q}=\left(\frac{\pi}{8a},0,0\right)$. The values of electric field is chosen so that the ratio $\omega_E/\bar{\omega}_B$ is rational (a,c,e,f) or irrational (b,d). (g,h) Semiclassical trajectory of the wave packet for different values of electric field when the magnetic flux is $\phi=0.002$ and the initial wave vector is $\bm{q}=\left(\frac{\pi}{30a},0,0\right)$. The electric field is chosen so that $\omega_E$ is commensurate with $\bar{\omega}_B\left(\frac{\pi}{8}\right)$, which slightly mismatches the cyclotron frequency. Other parameters are $\sigma=8a$, $N_x=N_y=201$, $N_z=501$.}
\end{figure}

The semiclassical trajectory of the electron is shown in Fig.~\ref{figSM:stark-cyclotron} for several cases of electric and magnetic fields. Here, as the cyclotron frequency $\bar{\omega}_B$ is energy dependent, we index it using the initial wave vector along the $x$ direction. In particular, with $\bm{q}=(q_x, q_y, q_z)$, the associated cyclotron frequency is determined by the energy $\mathcal{E}(\bm{q}) = -2\gamma\left[\cos(q_xa) + \cos(q_ya)\right]$ and is denoted by $\bar{\omega}_B(q_xa)$ as we choose $q_y=0$ for the cases in Fig.~\ref{figSM:stark-cyclotron}. When the two frequencies are commensurate, the 3D orbits are closed cylindrical Lissajous curves [Fig.~\ref{figSM:stark-cyclotron}(a,c,e,f)]. When the ratio $\frac{\omega_E}{\bar{\omega}_B}$ is irrational, the trajectory becomes open but always remains on the surface of a finite-size cylinder [Fig.~\ref{figSM:stark-cyclotron}(b,d)]. Similarly, if the Bloch frequency is commensurate with the cyclotron frequency at an energy $\mathcal{E}(\pi/8,0)$ but the actual cyclotron frequency is given by $\mathcal{E}(\pi/30, 0)$, the orbits also become open and the electron resides on a finite-size cylinder [Fig.~\ref{figSM:stark-cyclotron}(g,h)]. The height and radius of this cylinder are inversely proportional to the electric and magnetic field strengths, respectively. Hence, the stronger the two fields are, the smaller these classical orbits become.
\section{Wannier-Stark-Landau spectrum}
This section provides supplemental data on the spectrum of the WSL states in a cubic lattice subjected to parallel electric and magnetic fields. The tight-binding Hamiltonian describing this system is given by
\begin{align}
	\mathcal{H} = \sum_{r,k_x,k_y,z}c^{\dagger}_{rk_xk_yz}\Big[Vz - 2\gamma\cos(2\pi\phi r+k_y)\Big]&c_{rk_xk_yz} - \sum_{r,k_x,k_y,z}\Big[c^{\dagger}_{(r+1)k_xk_yz}\gamma c_{rxyz} + c^{\dagger}_{rk_xk_y(z+1)}\gamma_zc_{rk_xk_yz}+ h.c.\Big] \nonumber\\
	&- \sum_{k_x,k_y,z}\Big[c^{\dagger}_{1k_xk_yz}\gamma e^{iqk_x}c_{qxyz} + h.c.\Big],
\end{align}
where $r=1\dots q$ labels the atomic sites within the magnetic unit cell. Defining the vector
\begin{equation}
	\mathbf{C}_{k_xk_y} = \begin{pmatrix}
		c_{1k_xk_y(-N)} & c_{1k_xk_y(-N+1)} & c_{1k_xk_y(-N+2)} & \dots & c_{1k_xk_y(N-1)} & c_{1k_xk_yN} & c_{2k_xk_y(-N)} & \dots & c_{qk_xk_y(N-1)} & c_{qk_xk_yN}
	\end{pmatrix}^T
\end{equation}
with $N$ determined by $N_z=2N+1$, the Hamiltonian can be expressed as
\begin{equation}
	\mathcal{H} = \sum_{k_x,k_y}\mathbf{C}^{\dagger}_{k_xk_y}H(k_x,k_y)\mathbf{C}_{k_xk_y}.
\end{equation}
Here, the Hamiltonian $H(k_x,k_y)$ is a $qN_z\times qN_z$ matrix, which can be written as a tensor sum of two matrices
\begin{equation}
	H(k_x,k_y) = \mathbb{I}_q\otimes H_{\text{W}} + H_{\text{L}}(k_x,k_y)\otimes\mathbb{I}_{N_z}\label{SEq: tensor sum}
\end{equation}
with
\begin{equation}
	H_{\text{W}} = \begin{pmatrix}
		\ddots & \vdots & \vdots & \vdots & \vdots & \vdots & \ddots\\
		\dots & -2V & -\gamma_z & 0 & 0 & 0 & \dots\\
		\dots & -\gamma_z & -V & -\gamma_z & 0 & 0 & \dots\\
		\dots & 0 & -\gamma_z & 0 & -\gamma_z & 0 & \dots\\
		\dots & 0 & 0 & -\gamma_z & V & -\gamma_z & \dots\\
		\dots & 0 & 0 & 0 & -\gamma_z & 2V & \dots\\
		\ddots & \vdots & \vdots & \vdots & \vdots & \vdots & \ddots
	\end{pmatrix}\label{EqSM: WS Hamiltonian}
\end{equation}
and
\begin{align}
	&H_{\text{L}}(k_x,k_y) = \nonumber\\
	&\begin{pmatrix}
		-2\gamma\cos(2\pi\phi + k_y) & -\gamma & 0 & 0 & \dots & -\gamma e^{iqk_x}\\
		-\gamma & -2\gamma\cos(4\pi\phi + k_y) & -\gamma & 0 & \dots & 0\\
		0 & -\gamma & -2\gamma\cos(6\pi\phi + k_y) & -\gamma & \dots & 0\\
		0 & 0 & -\gamma & -2\gamma\cos(8\pi\phi + k_y) & \dots & 0\\
		\vdots & \vdots & \vdots & \ddots & \vdots & \vdots \\
		-\gamma e^{-iqk_x} & 0 & 0 & 0 & \dots & -2\gamma\cos(2\pi q\phi + k_y)
	\end{pmatrix}.\label{EqSM: Landau Hamiltonian}
\end{align}
As the Hamiltonian matrix is a tensor sum of two independent matrices, its spectrum can be straightforwardly obtained through the Minkowski sum
\begin{equation}
	E = \left\{\mathcal{E}_{\nu} + \mathcal{E}_n|\mathcal{E}_{\nu}\in E_{\text{W}},\mathcal{E}_n\in E_{\text{L}}\right\},
\end{equation} 
where $E_{\text{W}}$ and $E_{\text{L}}$ denote the spectra of $H_{\text{W}}$ and $H_{\text{L}}$, respectively.
\subsection{Bulk spectrum}
In the thermodynamic limit, the spectrum of Wannier-Stark Hamiltonian~Eq.~\eqref{EqSM: WS Hamiltonian} becomes $\mathcal{E}_{\nu}=V\nu$, as derived in the previous section. Consequently, the bulk spectrum of our cubic lattice is given by the Minkowski sum of $V\nu$ and the eigenvalues $\mathcal{E}_n(k_x,k_y)$ of $H_{\text{L}}(k_x,k_y)$ [Eq.~\eqref{EqSM: Landau Hamiltonian}]. We examine the spectrum in two cases.

\begin{figure}
	\includegraphics[width=\linewidth]{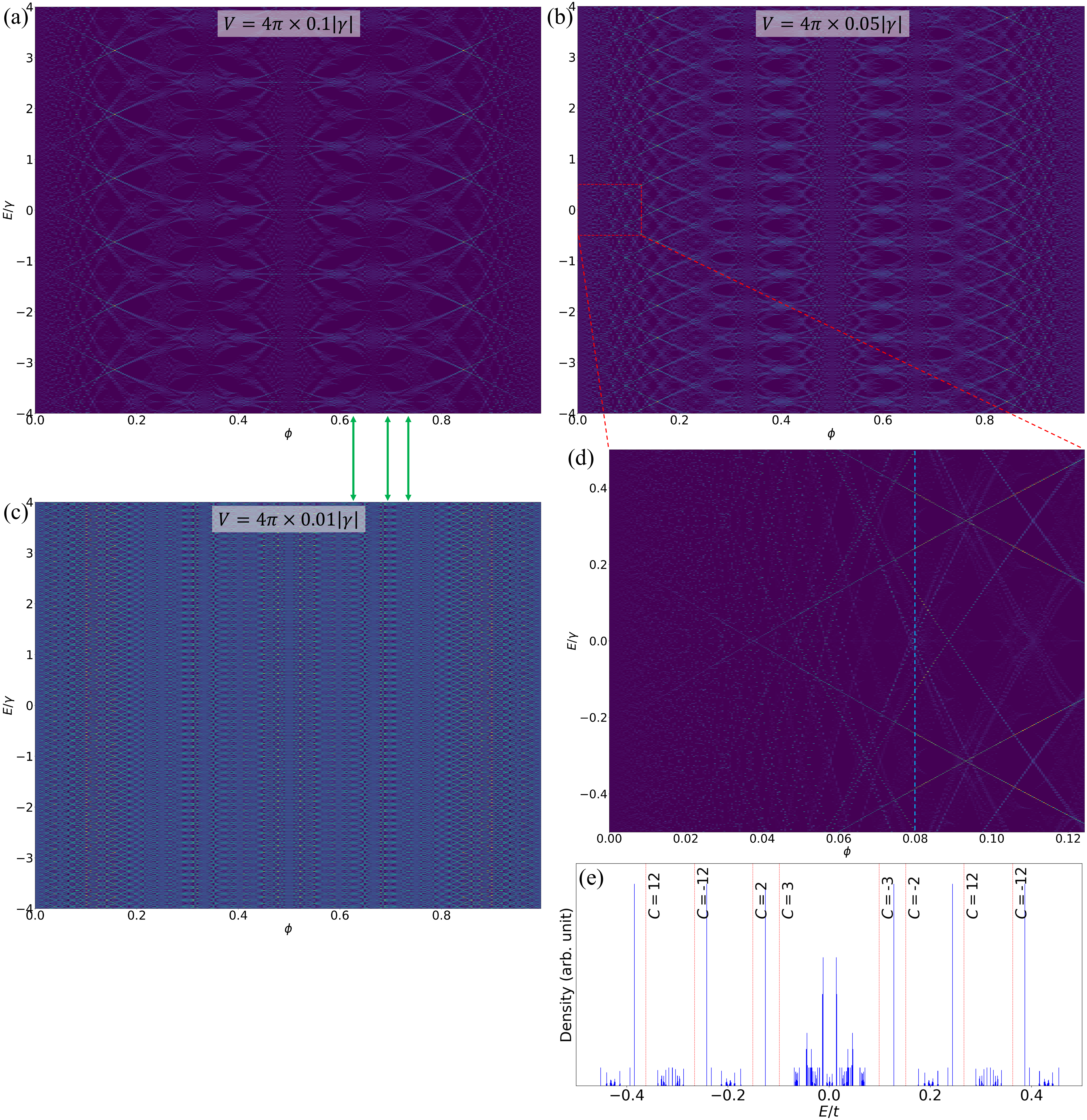}
	\caption{\label{figSM:spectra B} \textit{Spectral density of cubic lattice with varying magnetic field}. (a,b,c) The spectral density for three different values of the electric potential $V$. The electron's motion within the cubic lattice is described by the hopping parameters $\gamma=1$ and $\gamma_z=0.5$. The magnetic flux is rational, given by $\phi=p/q$ with $q=251$. At each value of $\phi$, the spectral density is pictured as a histogram of $N_p=1001$ boxes. (d) A zoom in of the area denoted by the red dashed box in (b). Here, we have $q=2003$ and $N_p=1000$. (e) The spectral density at a flux $\phi=\frac{11}{137}\approx0.08$, depicted by the blue dashed line in (d). The vertical red dotted lines mark some noticeable spectral gaps, accompanied by their Chern numbers.}
\end{figure}

First, fixing the electric potential at three different values, we obtain the Hofstadter-Stark spectra shown in Fig.~\ref{figSM:spectra B}. These spectra are a collection of replicas of the Hofstadter butterfly of a square lattice, shifted by $V\nu$, as can be seen clearly in Figs.~\ref{figSM:spectra B}(a,b) for a strong electric field. Consequently, the spectra are periodic in energy with a period $V$, analogous to the Floquet quasienergy spectrum of a periodically driven system. Due to this periodicity, we only need to show a region of these unbounded spectra; here we choose $E/\gamma$ to vary from $-4$ to $+4$ in Figs.~\ref{figSM:spectra B}(a,b,c). In a weak electric field [Fig.~\ref{figSM:spectra B}(c)], the spectrum is nearly gapless because the replicas of the Hofstadter butterfly closely overlap. Meanwhile, in a strong electric field [Figs.~\ref{figSM:spectra B}(a,b)], we can find spectral gaps within the spectrum, as demonstrated in Fig.~\ref{figSM:spectra B}(d,e). Most of these gaps are nontrivial as they have nonzero Chern numbers [Fig.~\ref{figSM:spectra B}(e)], which are a combination of the Chern numbers of spectral gaps in the Hofstadter butterflies. Here, the Chern numbers are computed using the Fukui-Hatsugai-Suzuki method~\cite{fukui2005chern}. Importantly, the Chern numbers of spectral gaps of the WSL states are finite due to the finite bandwidth of $E_{\text{L}}$, which is different from the continuum picture discussed in the main text. Additionally, in the weak field regime [Fig.~\ref{figSM:spectra B}(c)], there are several small ranges of $\phi$ where the spectrum exhibits equidistant resonances. Comparing the spectra of Figs.~\ref{figSM:spectra B}(a,c), we see that these equidistant resonances appear at magnetic fluxes where the square-lattice Hofstadter spectrum is highly fractal and features some locally horizontal resonances, as marked by the green arrows.

\begin{figure}
	\includegraphics[width=\linewidth]{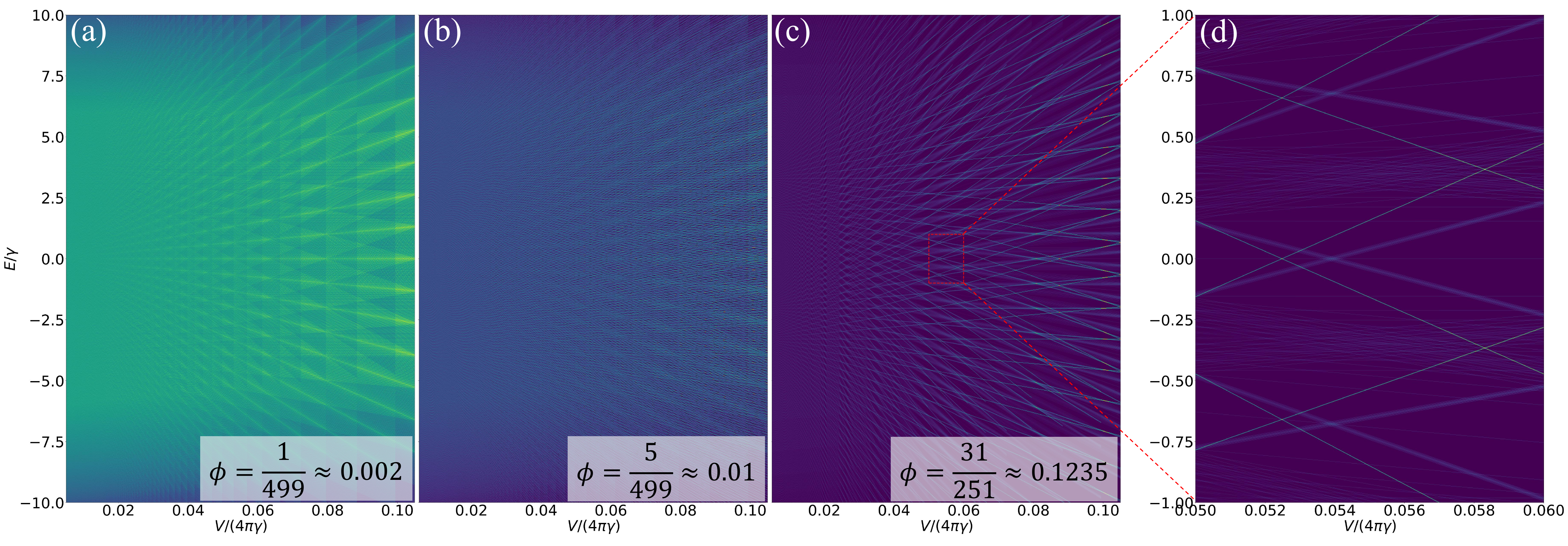}
	\caption{\label{figSM:spectra E} \textit{Spectral density of the cubic lattice with varying electric field}. (a,b,c) The spectral density for three different rational values of the magnetic flux $\phi$. The electron's motion within the cubic lattice is described by the hopping parameters $\gamma=1$ and $\gamma_z=0.5$. At each value of the electric potential $V$, the spectral density is pictured as a histogram of $N_p=1001$ boxes. (d) A zoom in of the area denoted by the red dashed box in (c).}
\end{figure}

Second, fixing the magnetic flux at three different values, we obtain the variation of the spectra against the electric potential $V$, which are shown in Fig.~\ref{figSM:spectra E}. The spectrum exhibits a fractal-like, filamentary structure with bright thread-like levels crossing at nodes, sitting over a darker diffuse background, with the network becoming denser for weaker fields [Figs.~\ref{figSM:spectra E}(a,b,c)]. However, by zooming into the spectrum [Fig.~\ref{figSM:spectra E}(d)], we see that the fractal structure is incomplete as the thread-like density distribution becomes sparser. This reduction in density stems from the finite bandwidth of the Landau spectrum $E_{\text{L}}$: Assuming that $E_{\text{L}}$ has a bandwidth of $W\le4|\gamma|$, only a finite number of sets $\left\{\mathcal{E}_{\nu} + \mathcal{E}_n||\mathcal{E}_{\nu}|\le W/2,\mathcal{E}_n\in E_{\text{L}}\right\}$ have levels that cross the zero energy point. This number decreases at greater values of $V$, resulting in a sparser spectral distribution and more visible energy gaps.
\subsection{Slab spectrum}
\begin{figure}
	\includegraphics[width=\linewidth]{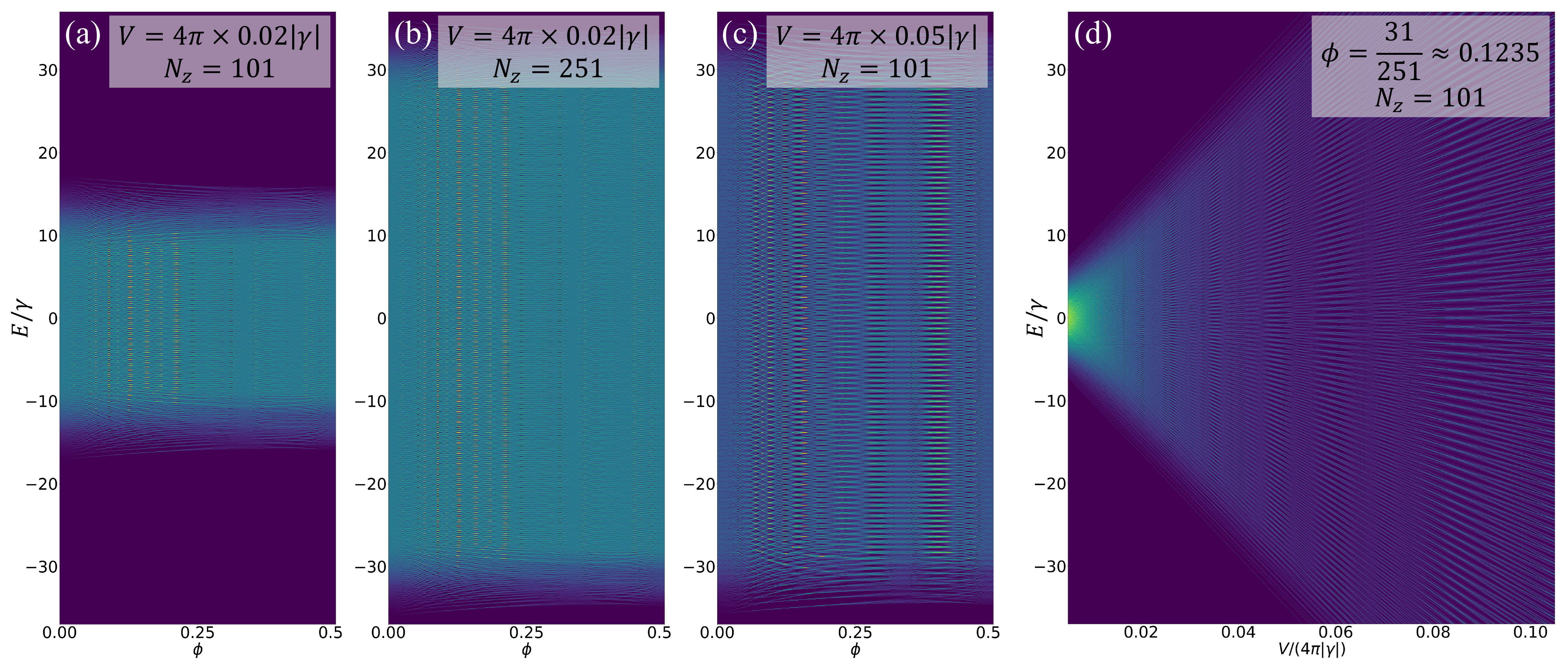}
	\caption{\label{figSM:spectra slab} \textit{Spectral density of the cubic lattice confined along the $z$ direction}. (a,b,c) The spectral density against the magnetic flux $\phi$ for different values of the electric potential $V$ and slab thickness $N_z$. At each value of the magnetic flux, the spectral density is pictured as a histogram of $N_p=1001$ boxes. (d) The spectral density against the electric potential $V$ for a fixed magnetic flux $\phi$ and slab thickness $N_z$.}
\end{figure}

In the main text, we considered the Hofstadter-Stark spectrum of a finite slab confined along the $z$ direction with $N_z$ layers - the geometry used to obtain the quantized Hall conductance and the chiral hinge modes. This differs from the bulk case of the previous subsection, where the Wannier-Stark ladder is unbounded: for a finite slab, there are only $N_z$ discrete Wannier-Stark levels, so the total spectral extent along $z$ is bounded and grows with both $V$ and $N_z$ [cf. Fig.~\ref{figSM:bloch}(e)]. Figure~\ref{figSM:spectra slab} illustrates this dependence in the spectrum of WSL states.

Panels (a) and (c) fix the slab thickness at $N_z = 101$ and compare two field strengths, $V = 4\pi\times0.02|\gamma|$ and $V = 4\pi\times0.05|\gamma|$. Increasing $V$ widens the overall spectral range, consistent with the wider-spaced Wannier-Stark ladder shown in Fig.~\ref{figSM:bloch}(e), while at the same time thinning out the density of overlapping levels - visible gaps open between the discrete equidistant resonances, mirroring the bulk behavior already seen in Fig.~\ref{figSM:spectra E}. Panels (a) and (b) instead fix $V = 4\pi\times0.02|\gamma|$ and compare two slab thicknesses, $N_z = 101$ and $N_z = 251$. Here, the spectrum widens for a related but distinct reason: a thicker slab hosts more Wannier-Stark levels, so their combined energy range grows even though the spacing between adjacent levels (set by $V$) is unchanged.

Panel (d) shows the complementary cut: at fixed flux $\phi = 31/251 \approx 0.1235$ and $N_z = 101$, the spectral density is plotted directly against $V$. As $V$ increases, the initially narrow, densely packed spectrum fans out and its density thins, in line with the same $V$-dependence of the Wannier-Stark spacing discussed above. Taken together, Fig.~\ref{figSM:spectra slab} shows that the slab spectrum interpolates between a gapless, quasi-continuum regime at weak fields or thin slabs, and a well-separated-resonances regime at strong fields - the latter being the regime relevant for the quantized Hall conductance discussed in the previous subsection.

\end{document}